\documentclass[10pt,aps,pra,twocolumn,nofootinbib,longbibliography,superscriptaddress,floatfix]{revtex4-2}
\usepackage{amsmath,amssymb}  
\usepackage{graphicx}
\usepackage{physics}
\usepackage{comment}
\usepackage[final]{hyperref} 
\hypersetup{
	colorlinks=true,      
	linkcolor=blue,     
	citecolor=magenta,      
	filecolor=magenta,  
	urlcolor=blue         
}
\begin{document}
\title{\large Dissipative quantum phase transitions monitored by current fluctuations}
\author{Masataka Matsumoto}
\email{masataka@sjtu.edu.cn}
\author{Zi Cai}
\author{Matteo Baggioli}
\email{b.matteo@sjtu.edu.cn}
\affiliation{Wilczek Quantum Center, School of Physics and Astronomy, Shanghai Jiao Tong University, Shanghai 200240, China}
\affiliation{Shanghai Research Center for Quantum Sciences, Shanghai 201315, China}
\begin{abstract}
Dissipative phase transitions (DPT) are defined by sudden changes in the physical properties of nonequilibrium open quantum systems and they present characteristics that have no analog in closed and thermal systems. Several methods to detect and characterize DPT have been suggested in the literature, the most famous of which -- the \textit{Liouvillian gap} -- can be derived from a spectral analysis of the Liouvillian super-operator that governs the complex interplay between coherent and dissipative dynamics. Here, we consider the \textit{output current}, defined as the average total quantum jumps per unit time between the open quantum system and the environment. We propose that output current fluctuations, and in particular their dynamical correlations, their power spectrum, and their characteristic timescale can provide valuable information about DPT, confirming a dramatic change of behavior at the critical point.
We validate our proposal using the dissipative XYZ model and the nonlinear driven-dissipative Kerr model, showing good agreement with previous estimates of the location of the critical point. Compared to previous approaches, our proposal could already be experimentally tested in optical systems, providing a practical method to detect criticality in quantum open systems.
\end{abstract}

\maketitle

\section*{Introduction} \label{sec:intro}
Open quantum many-body systems have gained significant attention in recent years, acquiring a  fundamental role in quantum physics and technological applications \cite{rivas2012open,breuer2002theory,rotter2015review}.
Unlike closed systems, which evolve under unitary dynamics, open quantum systems interact with an external environment and are governed by non-unitary dynamics due to their dissipative nature.
The interaction with the environment introduces a rich variety of phenomena, such as non-equilibrium steady states and non-equilibrium quantum phase transitions that go beyond the paradigm of equilibrium statistical physics \cite{schaller2014open}.

Dissipative phase transitions (DPT) are critical phenomena in open quantum systems that correspond to an abrupt change in their physical properties (see \textit{e.g.} \cite{PhysRevLett.105.015702,PhysRevLett.118.247402,Fitzpatrick2017,Fink2017prx,PhysRevX.5.031028,Kessler:2012wsp,PhysRevLett.101.105701,PhysRevLett.97.236808}), reminiscent of the physics of phase transitions in closed systems.
The competition between the coherent and incoherent dissipative effects can trigger the emergence of multiple steady states and transitions between them in the thermodynamic limit \cite{Kessler:2012wsp,Minganti:2018kgs}.
Dissipative phase transitions have been extensively studied in photonic systems \cite{Charmichael2015,Bartolo2016,Mendoza2016,Biella2017,Savona2017,FossFeig2017,Vicentini2018} and spin systems \cite{Kessler:2012wsp,Lee2013,Jin2016,Rota2017,rota2018dynamical,Li2022}; nevertheless, plenty of fundamental questions remain open.

An essential aspect of DPT concerns the question of how to accurately locate the associated critical points and how to define universal physical observables able to probe them. In this regard, the \textit{Liouvillian gap}, defined as the real part of the lowest eigenvalue of the Liouvillian super-operator, is considered the most efficient observable to probe the onset of criticality at the DPT. In particular, DPTs can be identified by the location at which the Liouvillian gap vanishes, reminiscent of the phenomenon of {\it critical slowing down} in second order phase transitions \cite{Kessler:2012wsp,Minganti:2018kgs}. Nevertheless, it has been shown that the Liouvillian gap is not enough to achieve a complete characterization of DPTs, motivating the definition of additional spectral gaps (\textit{e.g.}, \cite{PhysRevB.110.104303}). In general, other methods have been developed to characterize the critical behaviors of DPTs, including fidelity susceptibility \cite{Li2022}, cluster mean-field methods \cite{Jin2016}, trace distance susceptibility \cite{Li2022},  stochastic trajectory calculations \cite{rota2018dynamical}, quantum Fisher information and entanglement \cite{Rota2017}, and so on.

\begin{figure}[tbp]
    \centering
    \includegraphics[width=0.9\linewidth]{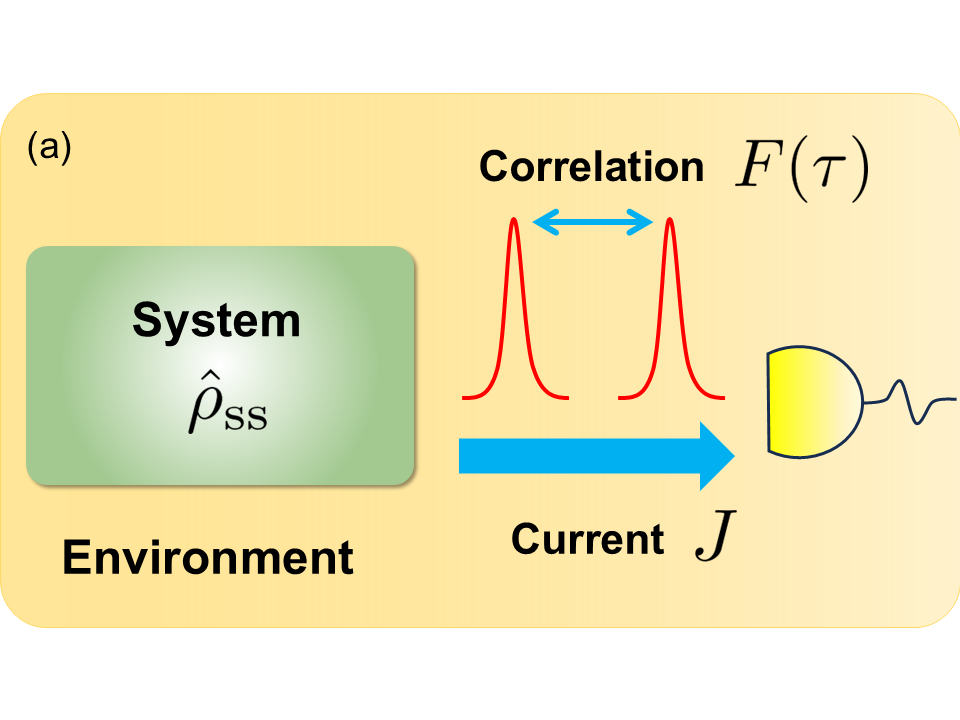}
    \includegraphics[width=0.9\linewidth]{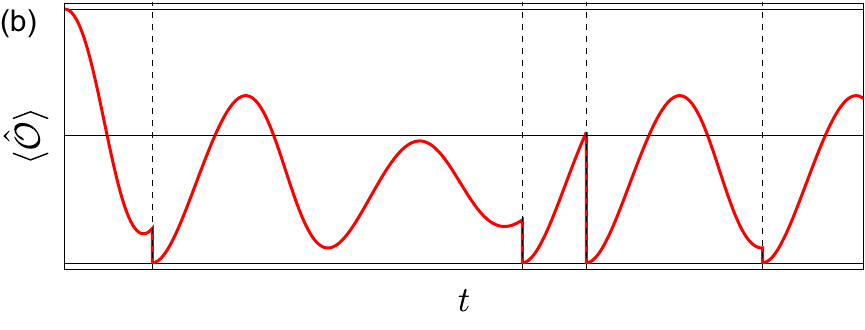}
    \caption{\textbf{(a)} Schematic picture of a typical open quantum system. The system is coupled to the environment, in which it dissipates incoherently. \textbf{(b)} Depiction of a quantum trajectory for an operator $\hat{{\cal{O}}}$. The quantum jumps in a stochastic process, denoted by gray dashed lines, are measured by the detector placed on the environment. The output current $J$ is defined by the average of the total quantum jumps per unit time.}
    \label{fig:system}
\end{figure}

In this work, we propose an alternative approach to locate and characterize DPTs that is based on the concept of output current fluctuations. The output current $J$ describes the average number of quantum jumps between the open quantum system and its coupled environment (see Fig.~\ref{fig:system} for a cartoon). Output current fluctuations contain important information about the open quantum system and its dynamics (see \cite{Landi:2023ktg} for a recent tutorial on the topic). Of particular significance is that the relationship between output current fluctuations and DPTs is experimentally accessible \cite{fink2018signatures,beaulieu2025observation}.

Here, we advance the idea that these fluctuations and in particular their time correlation $F(t)$, and the corresponding power spectrum $S(\omega)$, could serve as physical probes to characterize the critical point associated with the DPT, revealing a drastic transition in the dynamics. In particular, we show that output current fluctuations display an abrupt dynamical crossover between an overdamped to underdamped oscillating behavior that corresponds to the dissipative phase transition between two nonequilibrium steady states. 
We also consider the characteristic timescale of output current fluctuations and show that its slowing down universally locates the critical point as well.
The validity of our proposal is confirmed using two different benchmark models, the dissipative spin-1/2 XYZ Heisenberg model and the driven-dissipative Kerr model.
Finally, in the mean-field approximation, we provide a physical interpretation of our results in terms of Rabi frequency and the mean-field order parameter.

\section*{Methods} \label{sec:method}
We consider an open quantum system coupled to an environment whose time evolution is described by a quantum master equation (QME) \cite{PRXQuantum.5.020202},
\begin{equation}
    \frac{\dd \hat{\rho}}{\dd t} = {\cal L} \hat{\rho} = -i\left[ \hat{H}, \hat{\rho}\right] + \sum_{j} {\cal D}_{j}[\hat{\rho}],
    \label{eq:QME}
\end{equation}
where ${\cal{L}}$ is the Liouvillian super-operator, $\hat{H}$ is a Hermitian Hamiltonian, and $\hat{\rho}$ is the density matrix.
The last term in Eq.~\eqref{eq:QME}, referred to as the dissipator ${\cal{D}}_{j}$, represents the coupling to the external environment and is defined as
\begin{equation}
    {\cal{D}}_{j}[\hat{\rho}] = \hat{L}_{j} \hat{\rho} \hat{L}_{j}^{\dagger} - \frac{1}{2} \left\{ \hat{L}^{\dagger}_{j}\hat{L}_{j}, \hat{\rho} \right\}.
\end{equation}
Here, $L_{j}$ is the Lindblad operator acting on a site $j$.
The solutions of the QME \eqref{eq:QME} can be formally written as
\begin{equation}
    \hat{\rho}(t) = e^{{\cal{L}}t} \hat{\rho}_{0},
\end{equation}
in terms of the initial density matrix $\hat{\rho}_{0}$.
We assume that the system eventually relaxes to a steady state $\hat{\rho}_{\rm ss}$ such that ${\cal{L}}\hat{\rho}_{\rm ss} =0$.
This implies that the set of eigenvalues of the Liouvillian super-operator, $\{\lambda_{i}\}~(i=0,1,2,\cdots)$, contains at least one zero eigenvalue.
Note that, since the Liouvillian is a non-Hermitian super-operator, its eigenvalues are complex valued.
The real part of all eigenvalues can be proven to be negative~\cite{breuer2002theory,rivas2012open}, $\Re[\lambda_{i}] \leq 0$.
Sorting the eigenvalues by the amplitude of their real part, we write $0=\Re[\lambda_{0}]\geq \Re[\lambda_{1}]\geq \cdots$.
Here, the second eigenvalue $\lambda_{1}$ controls the relaxation rate of the system in the long-time limit and it is called the {\it Liouvillian gap} or {\it asymptotic decay rate} \cite{Kessler:2012wsp}.

At the onset of a dissipative phase transition, the Liouvillian gap vanishes, indicating the emergence of another non-trivial steady state \cite{Minganti:2018kgs}. More formally, in the thermodynamic limit, $N\to +\infty$, an $M$th-order dissipative phase transition can be defined by the mathematical condition
\begin{equation}
    \lim_{\zeta \to \zeta_{c}} \left| \frac{\partial^{M}}{\partial \zeta^{M}} \lim_{N\to +\infty} \hat{\expval{O}}_{\rm ss} \right| \to +\infty,
\end{equation}
where $\zeta_{c}$ is the critical value of the controlling parameter $\zeta$ and $\hat{\expval{O}}_{\rm ss} = \Tr[\hat{O}\hat{\rho}_{\rm ss}(\zeta,N)]$ is the expectation value of an observable $\hat{O}$ in the steady state.
This discontinuous behavior in $\hat{\rho}_{\rm ss}$ in the thermodynamic limit is associated with the closure of the Liouvillian gap and the emergence of multiple steady states \cite{Minganti:2018kgs}.

On the other hand, the output current $J(t)$ is emerging as an interesting physical quantity to characterize open quantum systems \cite{Landi:2023ktg}. Here $J(t)$ is defined by the average of the stochastic current $I(t)$, $J(t)\equiv \mathrm{E}[I(t)] $, where $\mathrm{E}[\cdot]$ denotes a classical average over stochastic trajectories.
The stochastic current $I(t)$ counts the net number of quantum jumps from the system to the environment in units of time, associated with the jump term in the dissipator $\hat{L}_{j} \hat{\rho}(t)\hat{L}_{j}^{\dagger}$.
More precisely, one finds
\begin{equation}
     J(t) = \sum_{j} \nu_{j} \Tr \left[ \hat{L}_{j} \hat{\rho}(t)\hat{L}_{j}^{\dagger} \right],
\end{equation}
where $\nu_{j}$ denotes the weights associated with the physical process in question.
In what follows, we choose $\nu_{j}=-1$, which is a natural choice if we consider only a dissipative process.
Finally, we notice that, if we consider a steady state $\hat{\rho}_{\rm ss}$, the output current is a constant number that does not evolve in time.

In a steady state, the time correlations between stochastic currents, $I(t)$ and $I(t+\tau)$, are quantified by the two-point correlation function,
\begin{align}
    F(t,t+\tau) &= \mathrm{E} \left[ \delta I(t) \delta I(t+\tau) \right] \nonumber\\
    &= \mathrm{E}\left[ I(t) I(t+\tau) \right] - J^{2},
\end{align}
where $\delta I(t) = I(t) - J$ is the current fluctuation with the output current $J=\mathrm{E}[I(t)]=\mathrm{E}[I(t+\tau)]$.
For simplicity, we denote the correlation function by $F(\tau)$, by choosing $t=0$.
For quantum jumps, $F(\tau)$ can be explicitly written as
\begin{equation}
    F(\tau) = K \delta(\tau) + \Tr \left[ {\cal{J}} e^{{\cal{L}} \abs{\tau}} {\cal{J} \hat{\rho}_{\rm ss}} \right] - J^{2},
    \label{eq:Ftau}
\end{equation}
where
\begin{equation}
    K = \sum_{j} \nu_{j}^{2} \Tr \left[ \hat{L}_{j} \hat{\rho}_{\rm ss} \hat{L}^{\dagger}_{j} \right],
\end{equation}
and ${\cal{J}}$ is the super-operator for the output current:
\begin{equation}
    {\cal{J}}\hat{\rho} = \sum_{j} \nu_{j} \hat{L}_{j} \hat{\rho} \hat{L}^{\dagger}_{j}.
\end{equation}
The first term in Eq. \eqref{eq:Ftau} corresponds to the contribution from uncorrelated white noise. For later convenience, we define the correlation function after subtracting the white noise contribution as
\begin{equation}
    C(\tau) \equiv F(\tau) - K \delta (\tau)= \Tr \left[ {\cal{J}} e^{{\cal{L}} \abs{\tau}} {\cal{J} \hat{\rho}_{\rm ss}} \right] - J^{2}.
    \label{eq:C}
\end{equation}

Performing the Fourier transform of $F(\tau)$ in Eq. \eqref{eq:Ftau}, we obtain the {\it power spectrum}:
\begin{equation}
    S(\omega) = \int_{-\infty}^{\infty} \dd \tau F(\tau)e^{-i\omega \tau}.
    \label{eq:Somega}
\end{equation}
Since the correlation function is an even function of $\tau$ in the steady state, $F(\tau) = F(-\tau)$, the power spectrum is real and even in frequency, $S(\omega) = S(\omega)^{*} = S(-\omega)$. 

\section*{Results} \label{sec:result}
\subsection{Dissipative XYZ model} \label{sec:XYZ}
We consider a two-dimensional spin-1/2 Heisenberg XYZ model, whose Hamiltonian is given by
\begin{equation}
    \hat{H} = \sum_{\left< i,j\right>} \left( 
    J_{x} \hat{\sigma}_{i}^{x} \hat{\sigma}_{j}^{x} +
    J_{y} \hat{\sigma}_{i}^{y} \hat{\sigma}_{j}^{y} +
    J_{z} \hat{\sigma}_{i}^{z} \hat{\sigma}_{j}^{z} \right),
\end{equation}
where $\hat{\sigma}_{i}^{\alpha}\,(\alpha=x,y,z)$ are the Pauli matrices on the $i$th-site and $J_{\alpha}$ is the coupling between nearest-neighbor spins.
The summation is over nearest neighbors in the two-dimensional lattice with periodic boundary conditions.
We assume that each spin is independently coupled to the environment, which tends to flip the spins down incoherently.
Thus, the dissipative part of the Liouvillian super-operator is given by
\begin{equation}
    \sum_{j} {\cal{D}}_{j}[\hat{\rho}] = \gamma \sum_{j} \left[ \hat{\sigma}_{j}^{-}\hat{\rho} \hat{\sigma}_{j}^{+} - \frac{1}{2}\{\hat{\sigma}_{j}^{+} \hat{\sigma}_{j}^{-}, \hat{\rho} \} \right],
\end{equation}
where $\gamma$ is the dissipation rate and $\hat{\sigma}^{\pm}_{j} = (\hat{\sigma}_{j}^{x} \pm i \hat{\sigma}_{j}^{y})/2$ are the raising and lowering operators.

When dissipation is strong enough, a steady state with all spins down along the $z$-axis is generally realized in the thermodynamic limit, which is referred to as the paramagnetic phase.
If, on the other hand, the anisotropy of the system is enhanced, a state that has non-zero expectation values of the magnetization along the $x$- and $y$-axes can emerge even in the thermodynamic limit.
This other ferromagnetic phase is defined by the spontaneous breaking of the original $Z_{2}$ symmetry.

First, let us briefly review the analysis of this model in the mean-field approximation.
In this limit, the density matrix of the total system can be factorized into a single-site density matrix, effectively reducing the many-body problem to a single-site system.
Thus, the mean-field Hamiltonian is given by
\begin{equation}
    \hat{H}_{\rm MF} = \sum_{\alpha} J_{\alpha}\expval{\hat{\sigma}^{\alpha} }\hat{\sigma}^{\alpha}, \label{eq:HMF}
\end{equation}
where $\expval{\hat{\sigma}^{\alpha} } = \Tr[\hat{\sigma}^{\alpha} \hat{\rho}]$.
Since the master equation in the mean-field approximation reduces to the nonlinear Bloch equation \cite{Lee2013}, the expectation values of $\hat{\sigma}^{\alpha}$ in the steady state can be derived analytically.
In this approximation, the magnetizations along the $xy$ plane, $\expval{\hat{\sigma}^{x,y}}$, are the order parameters which characterize the dissipative phase transition. The details of this analysis are discussed in Appendix \ref{app:A}.
In the mean-field approximation, we analytically obtain the critical value of $J_{y}$,
\begin{equation}
    J_{y}^{c} = J_{z} + \frac{\gamma^{2}}{256(J_{z}-J_{x})}. 
    \label{eq:Jyc}
\end{equation}
Choosing $J_{x}=0.9$ and $J_{z}=1.0$, the critical point is then located at $J_{y}^{c}\approx 1.04$ in the mean-field approximation.
For $J_{y}<J_{y}^{c}$, the system is in the paramagnetic phase with $\expval{\hat{\sigma}^{x,y}} =0$, whereas for $J_{y}>J_{y}^{c}$, it lies in the ferromagnetic phase with $\expval{\hat{\sigma}^{x,y}} \neq 0$.
For convenience, in the rest of paper, we work in units of $\gamma$ (or equivalently, we will set $\gamma=1$) unless explicitly stated otherwise.

We then proceed with computing the correlation function $C(\tau)$.
\begin{figure}[tbp]
    \centering
    \includegraphics[width=0.9\linewidth]{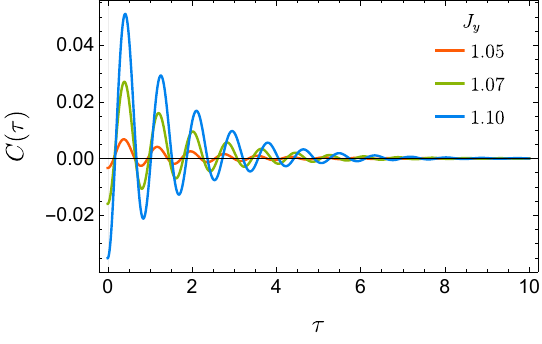}
    \caption{Correlation function $C(\tau)$ as a function of $\tau$ for several values of $J_{y}$ in the mean-field approximation.}
    \label{fig:FMF}
\end{figure}
Fig.~\ref{fig:FMF} shows the correlation function as a function of $\tau$ for several values of $J_{y}$ in the mean-field approximation. Only values of $J_y$ within the ferromagnetic phase are shown because the correlation function $C(\tau)$ identically vanishes in the paramagnetic phase in the mean-field approximation. Above the critical point, $J_{y}>J_{y}^{c}$, damped oscillations are observed. Importantly, the amplitude of these oscillations becomes smaller as the parameter $J_y$ approaches the critical point. Moreover, the damping rate of the oscillations decreases by approaching the critical point, indicating that the corresponding relaxation time becomes larger. Despite this being reminiscent of the phenomenon of critical slowing down, within the mean-field approximation, we do not observe any divergent time-scale at the critical point (see appendix \ref{app:A} for more details about this point and computations). Still, the results in Fig. \ref{fig:FMF} indicate that an oscillatory behavior in the correlation function $C(\tau)$ emerges associated with the dissipative phase transition. 

\begin{figure}[tbp]
    \centering
    \includegraphics[width=0.9\linewidth]{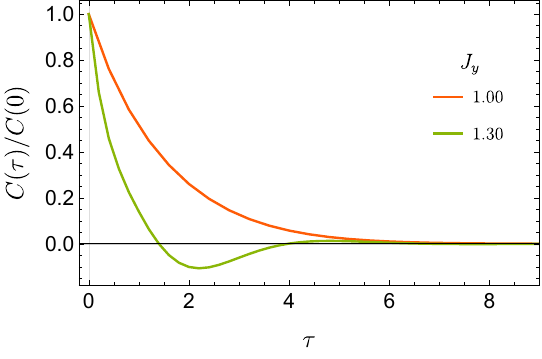}
    \caption{Normalized correlation function $C(\tau)/C(0)$ as a function of $\tau$ for two values of $J_{y}$. The results are for a $3\times 3$ square lattice with $J_{x}=0.9$ and $J_{z}=1$. The dynamical crossover between underdamped oscillating behavior and overdamped decaying behavior is observed.}
    \label{fig:Fplot}
\end{figure}

Going beyond the mean-field approximation, finite-size effects need to be carefully considered. It is known that finite-size systems do not exhibit phase transitions. On the other hand, as we will demonstrate, the dynamics of $C(\tau)$ will reveal a drastic change of behavior even in systems with small size.
Fig.~\ref{fig:Fplot} shows the correlation function, Eq. \eqref{eq:C}, for two values of $J_{y}$ in a $3\times 3$ lattice.
For small values of $J_y$, for example the orange line $J_y=1.00$, $C(\tau)$ displays a purely monotonic decay that, at late times, is well-approximated by a single exponential function. This corresponds to an overdamped response where correlations quickly decay. On the other hand, for large values of $J_y$, for example, the green line for $J_y=1.30$, the behavior of $C(\tau)$ changes dramatically since it is not purely monotonically decaying but it exhibits long-lived oscillations. This behavior corresponds to an underdamped response.

The separation between these two regimes marks the onset of a dynamical crossover.
From this observation, we consider the idea that a dissipative phase transition could be characterized by a dynamical crossover between overdamped and oscillating regimes in $C(\tau)$. We notice that this transition can be captured also in small finite size systems, since it is already observed in a $3\times 3$ lattice that is certainly far away from the thermodynamic limit.

In fact, the oscillations in $C(\tau)$, which are evident for large values of $J_y$, can be naturally interpreted as Rabi oscillations.
In the regime where dissipation dominates, the system is mostly pinned into the state with all spins down in the $z$-direction and the correlation of the current fluctuations monotonically decreases in time.
If the coherent effects in the system become larger, on the other hand, this leads to oscillations between states with positive and negative values of $\expval{\hat{\sigma}^{x,y}}$ in a quantum trajectory \cite{rota2018dynamical}, inducing Rabi oscillations.
Within the mean-field approximation, one can derive that a term of the form
\begin{equation}
J_{x}\expval{\hat{\sigma}^{x}}\hat{\sigma}^{x} \equiv \Omega_{\rm R} \hat{\sigma}^{x}\label{yuyu}
\end{equation}
induces Rabi oscillations with frequency $\Omega_{\rm R}$ that ultimately compete with the dissipative effects.
In other words, as explicitly shown in Eq. \eqref{yuyu}, a nonzero value of the order parameter $\expval{\hat{\sigma}^{x}}\neq 0$ leads to a finite Rabi frequency $\Omega_{\rm R}\neq 0$. A similar relaxation dynamics in other open quantum spin systems was discussed in \cite{Cai2013,Ren2020}.

\begin{figure}[tbp]
    \centering
    \includegraphics[width=0.9\linewidth]{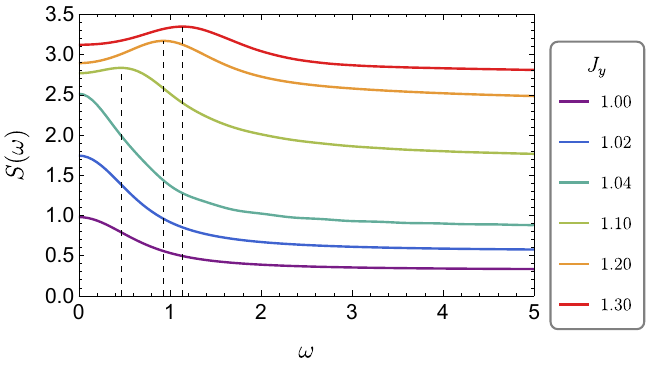}
    \caption{Power spectrum $S(\omega)$ for several values of $J_{y}$. 
    The black dashed lines denote the peak position for each curve.
    The results are for a $3\times 3$ square lattice with $J_{x}=0.9$ and $J_{z}=1$.}
    \label{fig:Splot}
\end{figure}

In order to study these dynamics in more detail, we compute the power spectrum, Eq. \eqref{eq:Somega}. To analyze its characteristics, we assume a damped harmonic oscillator (DHO) form,
\begin{eqnarray}\label{power}
    S(\omega) &\propto& \frac{1}{(\omega^{2}-\omega_0^{2})^{2}+\gamma_{0}^{2}\omega^{2}} \nonumber \\
    &=& \frac{1}{(\omega^{2}-\left(\Omega_{\rm R}^{2} + \gamma_{0}^{2}/4 \right))^{2}+\gamma_{0}^{2}\omega^{2}},
\end{eqnarray}
where $\omega_{0}$ and $\gamma_{0}$ denote the natural frequency and damping rate, respectively. Here, we identify the real part of the frequency as the Rabi frequency
\begin{equation}
    \Omega_{\rm R} =\Re \left[ \sqrt{\omega_0^2 -\frac{\gamma_{0}^2}{4}} \right].
\end{equation}
We also notice that the power spectrum displays a peak at
\begin{equation}
    \omega_{\rm peak} = \sqrt{\omega_0^{2} - \frac{\gamma_{0}^{2}}{2} } = \sqrt{\Omega_{\rm R}^{2} - \frac{\gamma_{0}^{2}}{4} }. \label{eq:omegapeak}
\end{equation}
Therefore, we can observe a peak at finite frequency in the power spectrum when $\Omega_{\rm R}>\gamma_{0}/{2}$.
Importantly, we should emphasize that a finite order parameter $\Omega_{\rm R} > 0$ does not immediately lead to the emergence of the finite $\omega_{\rm peak}$.
The transition between oscillating to decaying behavior in $C(\tau)$ does not exactly coincide with the shift of the peak position from $\omega=0$ to $\omega\neq 0$ in $S(\omega)$. On the contrary, it corresponds to $\Omega_{\rm R} \neq 0$ in Eq.~\eqref{power}, which is equivalent to the onset of the DPT.

In any case, following our previous observations, the location of the critical point can be predicted by determining $\Omega_{\rm R}$ directly from the structure of the power spectrum $S(\omega)$. Fig.~\ref{fig:Splot} shows the power spectrum $S(\omega)$ for several values of $J_{y}$ across the critical point.
For small values of $J_y$, we find that $S(\omega)$ has a maximum at $\omega=0$ and then decays monotonically. 
For larger values of $J_y$, where $C(\tau)$ displays oscillating behavior, a finite frequency peak emerges in the power spectrum. Its position moves to higher frequency by increasing $J_y$. 

\begin{figure}[tb]
    \centering
    \includegraphics[width=0.9\linewidth]{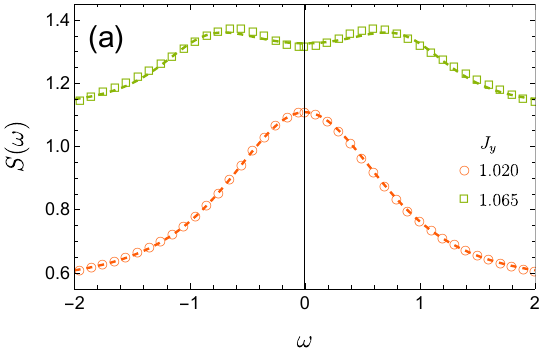}
    \includegraphics[width=0.85\linewidth]{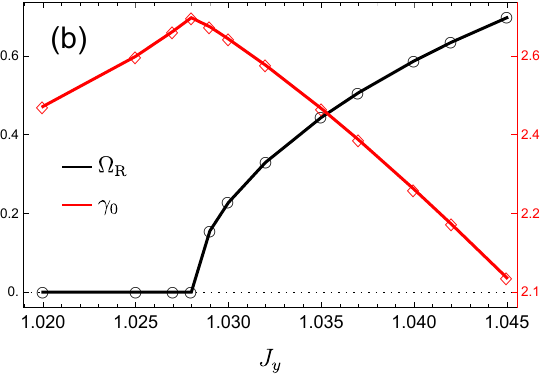}
    \caption{(\textbf{a}) Power spectrum $S(\omega)$ for two values of $J_{y}$ with the $2\times 3$ lattice. The dashed lines denote the fitting results with Eq.~(\ref{power}). (\textbf{b}) Fitting parameters $\{ \Omega_{\rm R}, \gamma_{0}\}$ as a function of $J_{y}$.}
    \label{fig:SwFit}
\end{figure}

By fitting $S(\omega)$ with Eq.~\eqref{power}
, we can directly extract $\Omega_{\rm R}$, $\gamma_{0}$, and $\omega_{\rm peak}$ as a function of $J_y$.
As an example, panel (a) of Fig.~\ref{fig:SwFit} shows the numerical plots of $S(\omega)$ and fitting results (dashed curves) for several values of $J_{y}$ in the $2 \times 3$ lattice.
We confirm that the numerical data around the critical point are well described by Eq.~(\ref{power}).
In panel (b) of Fig.~\ref{fig:SwFit}, we show the fitting parameters $ \Omega_{\rm R}$ (black) and $\gamma_{0}$ (red) as a function of $J_{y}$.
Although the peak position in $S(\omega)$ is at $\omega=0$ for the parameter region shown in panel (b) of Fig.~\ref{fig:SwFit}, the order parameter becomes finite for $J_{y} \gtrapprox 1.028$, corresponding to the onset of the oscillation in $F(\tau)$, as discussed above. 
The friction parameter $\gamma_{0}$ is gradually increased below the critical point but decreased above. Interestingly, it displays a maximum at the critical point where $\Omega_R \neq 0$.

\begin{figure}[tbp]
    \centering
    \includegraphics[width=0.9\linewidth]{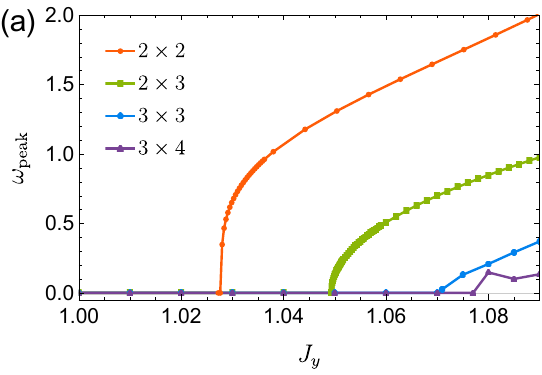}
    \includegraphics[width=0.9\linewidth]{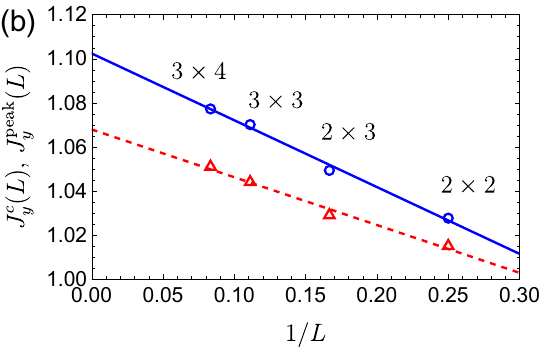}
    \caption{\textbf{(a)} Position of the peak in the power spectrum, $\omega_{\rm peak}$, as a function of $J_{y}$ with different lattice sizes $L$. \textbf{(b)} Finite-size scaling analysis of the critical parameter $J^c_y$ (\color{red}\textbf{red}\color{black}) and $J_{y}^{\rm peak}$ (\color{blue}\textbf{blue}\color{black}). In the thermodynamic limit, our method predicts a critical value of $J^{\rm c}_y \approx 1.065$, in very good agreement with other methods \cite{Jin2016,Rota2017,rota2018dynamical,PhysRevB.97.035103,Li2022}.}
    \label{fig:peakplot}
\end{figure}
In panel (a) of Fig.~\ref{fig:peakplot}, we show the peak position in the power spectrum, $\omega_{\rm peak}$, as a function of $J_{y}$. We present numerical results for different lattice sizes from $2\times 2$ to $3 \times 4$. In all cases, we observe that $\omega_{\rm peak}$ becomes nonzero above a specific value of $J_y$, defined as $J^{\rm peak}_y$. This value relates to (but does not coincide with) the location of the critical point $J^c_y$ that is defined by the value of $J_y$ at which $\Omega_{\rm R} \neq 0$.

The values of $J_{y}^{c}$ and $J^{\rm peak}_y$ are affected by finite-size effects and slightly grow with the size of the lattice. In panel (b) of Fig.~\ref{fig:peakplot}, we provide a linear fitting of $J^{\rm peak}_y$ and $J^c_y$ as a function of $1/L$. In general, we find that $J^{\rm peak}_y>J^c_y$, as one might expect. This implies that the position of the peak always overestimates the location of the dissipative phase transition. In the thermodynamic limit, $L \rightarrow \infty$, our method predicts $J^{\rm peak}_y \approx 1.10$ and $J^{\rm c}_y \approx 1.067$. 
The onset value of the peak shift is compatible with the mean-field result, $ J_{y}^{\rm peak} \approx 1.09$ (see Appendix \ref{app:A}).  
The location of the critical point, which is estimated via $J^c_y$, is also compatible with previous estimates from cluster mean-field methods \cite{Jin2016} ($J^c_y \approx 1.04$), quantum Fisher information with the corner-space renormalization method \cite{Rota2017} ($J^c_y \approx 1.07$), stochastic quantum trajectory calculations \cite{rota2018dynamical} ($J^c_y \approx 1.05$), linked-cluster expansions \cite{PhysRevB.97.035103} ($J^c_y \approx 1.0665$) and fidelity susceptibility \cite{Li2022} ($J^c_y \approx 1.05$).

We anticipate that this first approach based on a dynamical underdamped-overdamped crossover is very efficient in the XYZ model, but will become less accurate in more complex systems such as the driven-dissipative Kerr model (more details below). Because of this limitation, we consider a second approach, still based on output current fluctuations, that will work efficiently in both models considered. 

More in detail, we define the characteristic timescale of output current fluctuations by integrating the normalized autocorrelation function,
\begin{equation}
    \tau_{\rm s} \equiv \frac{1}{C(0)}\int_{0}^{\infty} \dd \tau ~C(\tau).\label{diff}
\end{equation}
We comment that $\tau_{\rm s}$ is directly related to the {\it quantum diffusion coefficient} defined as $D\equiv S(0)$ via the simple relation
\begin{equation}
    2 C(0) \tau_{\rm s} = D-K,
\end{equation}
with $K$ the strength of white noise fluctuations (see Eq. \eqref{eq:C}). The relation between the diffusion coefficient and DPTs was studied in \cite{kewming2022diverging}; more details about their analysis will be provided in the next section.

\begin{figure}[tbp]
    \centering
    \includegraphics[width=0.9\linewidth]{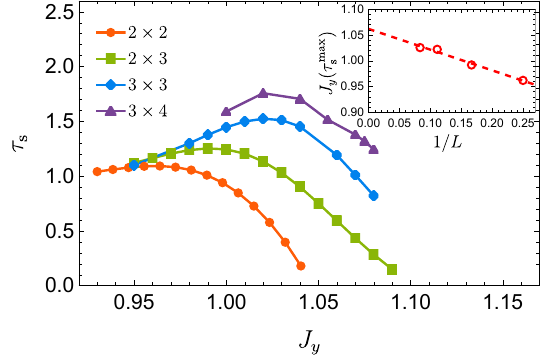}
    \caption{The plot of $\tau_{\rm s}$ as a function of $J_{y}$. The inset shows the value of $J_{y}$ at which $\tau_{\rm s}$ becomes maximum versus the inverse of the system size together with a linear fitting (dashed line).}
    \label{fig:DGplotX}
\end{figure}

Fig.~\ref{fig:DGplotX} shows the behavior of $\tau_{\rm s}$ as a function of $J_{y}$.
We find that $\tau_{\rm s}$ has a maximum at a certain value of $J_{y}$, whose intensity grows with the system size. 
This tendency resonates with the concept of \textit{critical slowing down}, which is conventionally associated with critical points in classical phase transitions but also in DPTs (\textit{e.g.}, a vanishing Liouvillian gap). Thus, we expect the peak in $\tau_{\rm s}(J_{y})$ to diverge in the thermodynamic limit and its location to be a good estimate for the critical point. In fact, as shown in the inset of Fig.~\ref{fig:DGplotX}, the value of $J_{y}$ at which $\tau_{\rm s}$ becomes maximum approaches $J_{y}^{c} \approx 1.062$ in the thermodynamic limit, in good agreement with other methods.

In the dissipative XYZ model, our results confirm that the characteristic frequency of the mode governing the late time output current fluctuations is a good observable to determine the onset of the DPT, which appears to be ultimately related to a dynamical underdamped-overdamped crossover in the real time correlators. We note that this criterion appears very natural within the mean-field approximation since it corresponds exactly to the emergence of a finite Rabi frequency and to a nonzero mean-field order parameter. Additionally, the critical point of the DPT is correctly predicted by the divergent behavior of the characteristic timescale $\tau_{\rm s}$, revealing further analogies with the physics of classical phase transitions, \textit{i.e.}, critical slowing down.

\subsection{Driven-Dissipative Kerr model} \label{sec:Kerr}
In order to confirm the validity of our new approach to DPTs based on output current fluctuations, we consider a second system known to display a dissipative phase transition: the driven-dissipative Kerr model.
For simplicity, we restrict ourselves to considering a typical nonlinear oscillator model with two-photon parametric driving and single-photon dissipation. In the case of two-photon driving the system exhibits a continuous phase transition characterized by the breaking of parity symmetry \cite{Bartolo2016,Minganti:2018kgs}.
The Hamiltonian in the rotating frame is given by
\begin{equation}
    H = -\Delta \hat{a}^{\dagger} \hat{a} + \frac{U}{2} \hat{a}^{\dagger} \hat{a}^{\dagger} \hat{a} \hat{a} + \frac{G}{4}\left( \hat{a}^{\dagger} \hat{a}^{\dagger} + \hat{a} \hat{a}  \right),
\end{equation}
where $\Delta$ is the detuning of the pumping and cavity frequency, $U$ is the coupling constant, and $G$ is the amplitude of the two-photon parametric driving.
The single-photon dissipation has the following form:
\begin{equation}
    {\cal{D}}[\hat{\rho}] = \gamma \left[ \hat{a} \hat{\rho} \hat{a}^{\dagger} - \frac{1}{2}\left\{ \hat{a}^{\dagger}\hat{a}, \hat{\rho} \right\}\right].
\end{equation}
The model has been analytically investigated by using the Keldysh formalism, as well as the mean-field approximation \cite{Zhang2021}.
We consider the resonant case $\Delta =0$ since we are interested in the continuous phase transition associated with the $Z_{2}$ symmetry breaking.
In the mean-field approximation, the coherent field amplitude $\alpha = \expval{a}$, considered as the order parameter, obeys the dynamical equation,
\begin{equation}
    \frac{\dd}{\dd t}\alpha = \left(-iU\abs{\alpha}^{2} - \frac{\gamma}{2} \right)\alpha - i\frac{G}{2} \alpha^{*}.
\end{equation}
In the steady state, we obtain the solution for the occupation number $n = \abs{\alpha}^{2}$ as
\begin{equation}
    n = 
    \begin{cases}
        0, & (G<\gamma) \\
         \sqrt{G^{2}-\gamma^{2}}/(2U), & (G>\gamma)
    \end{cases}
    \label{eq:KerrMF}
\end{equation}
where we assume that $G$ is a positive real number.
The thermodynamic limit can be achieved by taking $U\to 0$ because the numbers of photons, $n$, in the cavity is of order $\gamma/U$.
Therefore, the mean-field theory predicts the location of the critical point at $G_{c} =\gamma$. In the following, we always take $\gamma=1$.

\begin{figure}[tbp]
    \centering
    \includegraphics[width=0.9\linewidth]{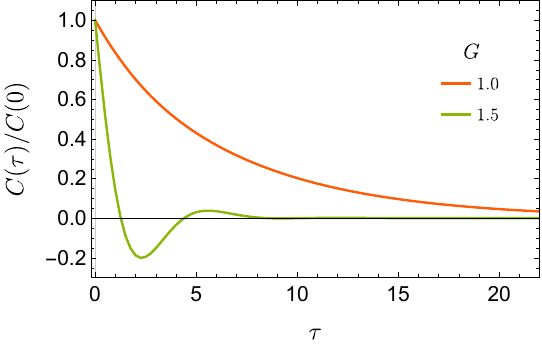}
    \caption{Correlation function $C(\tau)$, normalized by $C(0)$, for several values of $G$ with $U=1/30$ in the driven-dissipative Kerr model.}
    \label{fig:FplotK}
\end{figure}

In Fig.~\ref{fig:FplotK} we show the correlation function $C(\tau)$ for different values of the amplitude $G$ and fixed coupling constant $U$. For small $G$, the correlation function shows a overdamped behavior with a monotonic decay. On the other hand, for large $G$, a short time oscillatory behavior is apparent, indicating the onset of a dynamical crossover. 
 
The oscillatory behavior can be understood by the emergence of the effective detuning frequency $\Delta_{\rm eff}$, which is proportional to the order parameter, as analyzed in the power spectrum of the output current by the Keldysh formalism \cite{Zhang2021}. 

\begin{figure}[tbp]
    \centering
    \includegraphics[width=0.9\linewidth]{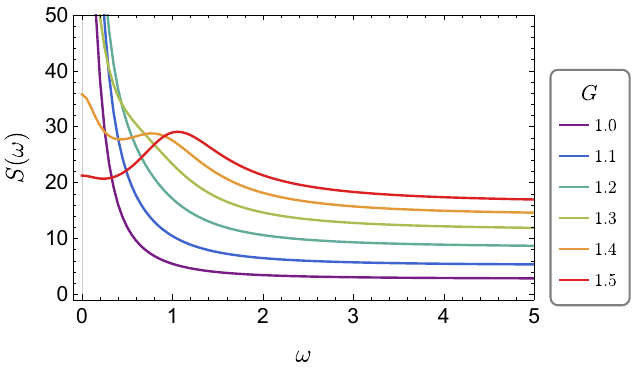}
    \caption{Power spectrum $S(\omega)$ for several values of $G$ with $U=1/30$ in the driven-dissipative Kerr model.}
    \label{fig:SplotK}
\end{figure}

We have repeated the same analysis of the output current power spectrum $S(\omega)$ shown in Fig.~\ref{fig:SplotK}, as done for the dissipative XYZ model in the preceding section. 
However, we find that a DHO power spectrum, Eq. \eqref{power}, does not fit the simulation data for the driven-dissipative Kerr model. 
This is because the pole structure of this model is more complex than a single pair of oscillators as it displays several overdamped (purely imaginary) modes coexisting with the damped oscillating mode related to the dynamics of the order parameter.
As a direct consequence of this more complex structure, the power spectrum in Fig.~\ref{fig:SplotK} exhibits two distinct peaks for larger $G$; the peak at the origin corresponds to the overdamped modes, while the peak at a finite $\omega$ corresponds to the damped oscillatory modes.

\begin{figure}[t]
    \centering
    \includegraphics[width=0.9\linewidth]{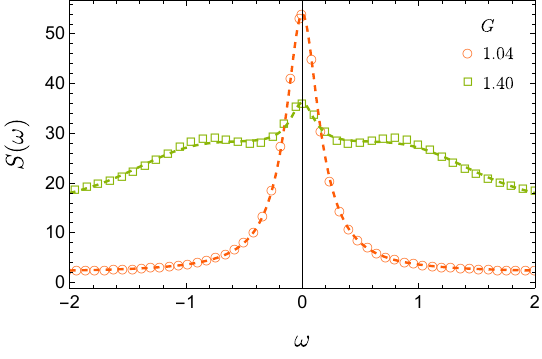}
    \caption{Power spectrum $S(\omega)$ for two values of $G$ with $U=1/30$. The dashed lines denote the fitting results with Eq.~(\ref{eq:powerK}).}
    \label{fig:SwfitK}
\end{figure}

\begin{figure}[t]
    \centering
    \includegraphics[width=0.9\linewidth]{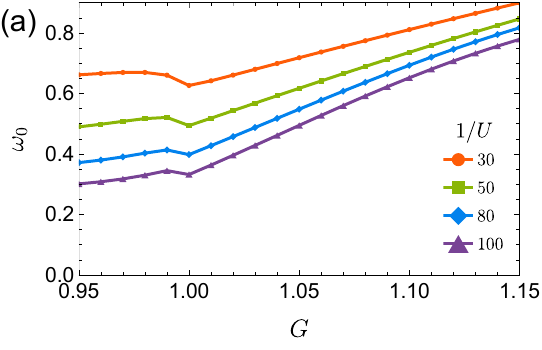}
    \includegraphics[width=0.9\linewidth]{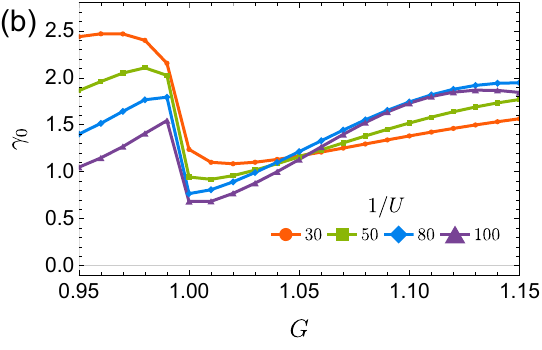}
    \includegraphics[width=0.9\linewidth]{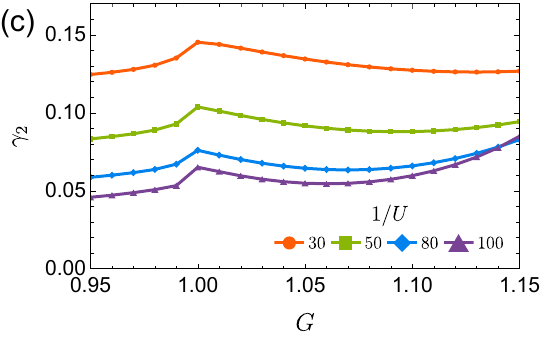}
    \caption{Fitting parameters (\textbf{a}) $\omega_{0}$, (\textbf{b}) $\gamma_{0}$, and (\textbf{c}) $\gamma_{2}$ as a function of $G$ for several values of $U$.}
    \label{fig:FitParaK}
\end{figure}

Now we assume that the power spectrum displays a more complex structure of the form
\begin{equation}
    S(\omega)\sim \frac{A_1}{(\omega^{2}-\omega_0^{2})^{2}+\gamma_{0}^{2}\omega^{2}} + \frac{A_2}{\omega^{2}+\gamma_{2}^{2}},
    \label{eq:powerK}
\end{equation}
where $\gamma_{2}$ is the decay rate of the purely damped mode and the parameters, $A_{1}$ and $A_{2}$, are determined by the fitting.
We confirm that our numerical results are well described by this form in a wide range of $G$ as shown in Fig.~\ref{fig:SwfitK}.
From the fitting, we show the fitting parameters as a function of $G$ for several values of $U$ in Fig.~\ref{fig:FitParaK}.
As shown, each fitting parameter shows a non-analytic behavior around the critical point, $G \approx 1$. However, it is difficult to predict the critical point efficiently, as in the XYZ model, because the additional fitting parameters reduce the accuracy of the method.

\begin{figure}[t]
    \centering
    \includegraphics[width=0.9\linewidth]{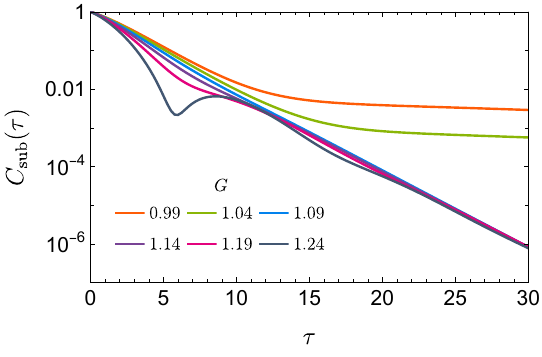}
    \caption{Correlation function (\ref{eq:Csub}) in which the overdamped mode at late time is subtracted for several values of $G$ with $U=1/30$.}
    \label{fig:Csub}
\end{figure}

Nonetheless, we assume that the connection between the onset of the oscillations in the correlation function and the dissipative phase transition could still hold even in the driven-dissipative Kerr model.
To prove this, we write the correlation function in real time as
\begin{equation}
    \frac{C(\tau)}{C(0)} = e^{-\gamma_{2} \tau} \left( 1 + e^{-\lambda \tau}+ \cdots \right),
\end{equation}
where $\gamma_{2}$ is the decay rate of the overdamped mode at late time and $\lambda$ can be complex. Then, we subtract the overdamped mode and define
\begin{equation}
    C_{\rm sub}(\tau) \equiv \frac{C(\tau)}{C(0) e^{-\gamma_{2}\tau}}. \label{eq:Csub}
\end{equation}
The value of $\gamma_{2}$ is obtained numerically by fitting of the late-time decay in $C(\tau)/C(0)$.
In Fig.~\ref{fig:Csub}, we show the subtracted correlation function $C_{\rm sub}(\tau)$ in logarithmic scale for several values of $G$ with $U=1/30$. 
For small $G$ ($G\lessapprox 1.04$), $C_{\rm sub}(\tau)$ approaches a constant late-time value in logarithmic scale, implying that is the overdamped mode with $\gamma_{2}$ dominating the behavior of $C(\tau)/C(0)$ at late time. For larger $G$, on the other hand, $C_{\rm sub}(\tau)$ shows an exponentially decaying behavior at late times, indicating the presence of an additional mode. As shown in the plot with $G=1.24$, for instance, the mode has a finite real frequency, inducing an oscillatory behavior in the real-time response. Compared to the plot of $C(\tau)/C(0)$, in this case, the oscillating behavior becomes easier to observe after subtracting the latest overdamped mode. As discussed above, our observations imply that the onset of the oscillations associated with the dissipative phase transition is hidden by the presence of additional overdamped modes that dominate the late-time response. In the driven-dissipative Kerr model, we find that at least one such mode exist, corresponding to the decay rate $\gamma_2$.  We acknowledge that this problem leads to a technical difficulty to precisely determine the critical point with our method in the driven-dissipative Kerr model. We also notice that the transition in Fig. \ref{fig:Csub} does not necessarily correspond to the dynamical crossover because it is still difficult to distinguish between underdamped and overdamped behaviors in $C_{\rm sub}(\tau)$.

\begin{figure}[tbp]
    \centering
    \includegraphics[width=0.9\linewidth]{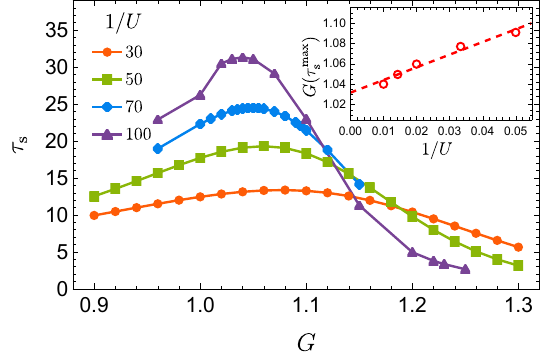}
    \caption{Plot of $\tau_{\rm s}$ as a function of $G$. The inset shows the value of $G$ at which $\tau_{\rm s}$ becomes maximum versus $1/U$ with a linear fitting.}
    \label{fig:DGplotK}
\end{figure}

Given the difficulties with predicting the critical point by looking at the finite frequency or finite time correlations, we move to the study of the characteristic timescale $\tau_{\rm s}$, as defined in Eq.~\eqref{diff}. In Fig.~\ref{fig:DGplotK}, we plot $\tau_{\rm s}$ as a function of $G$ for several values of $1/U$.
We confirm that $\tau_{\rm s}$ has a maximum at a certain value of $G$, as already observed in the dissipative XYZ model. We therefore utilize the position of such a peak as a proxy for the critical point.
The inset of Fig.~\ref{fig:DGplotK} shows our prediction for the location of the critical point together with a finite-size-scaling analysis. In the thermodynamic limit, $\tau_{\rm s}$ predicts a DPT at a value of $G_{c}\approx 1.03$, which is close to the mean-field result $G_{c}= 1$ and the previous prediction obtained using the fidelity susceptibility ($G_{c}\approx 1.04$) \cite{Li2022}.

We note that the diffusion coefficient $D$ has been already studied in the driven-dissipative Kerr model in \cite{kewming2022diverging}. In particular, its divergence near the continuous transition has been observed, but not analyzed in relation to the prediction of the critical point. By using the location of the maximum in $D$ as a criterion to locate the critical point, we find that this method consistently overestimates its position. This suggests that the correct quantity to use for this prediction is the characteristic timescale in Eq.~\eqref{diff}, rather than the quantum diffusion constant, further corroborating our proposal.

\subsubsection*{First-order phase transitions}
So far, we have focused on the second order dissipative phase transitions and the dynamical properties at the critical points. Here, in order to investigate the universality of our findings, we also consider a first-order dissipative phase transition.

\begin{figure}[tb]
    \centering
    \includegraphics[width=0.8\linewidth]{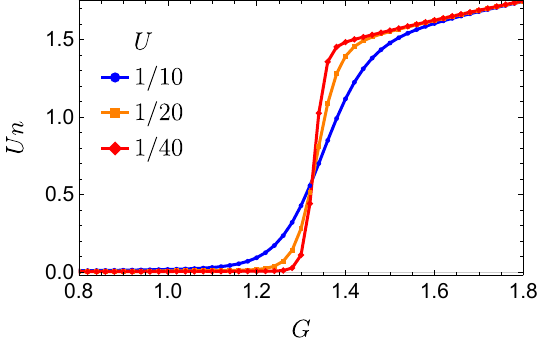}
    \caption{Occupation number $U n$ in the Kerr model as a function of $G$ for several values of $U$ with $\Delta =1$.}
    \label{fig:KerrOP1st}
\end{figure}

We employ the driven-dissipative Kerr model with finite detuning $\Delta\neq 0$. In Fig.~\ref{fig:KerrOP1st}, we show the occupation number $n=\abs{\alpha}^{2}$ as a function of $G$ for several values of $U$ with $\Delta=1$. For smaller $U$, one can see that it sharply changes at a certain value of $G$, implying the first-order phase transition in the thermodynamic limit ($U\to 0$). For $\Delta=1$, this value is approximately $G \approx 1.33$.

\begin{figure}[tb]
    \centering
    \includegraphics[width=0.8\linewidth]{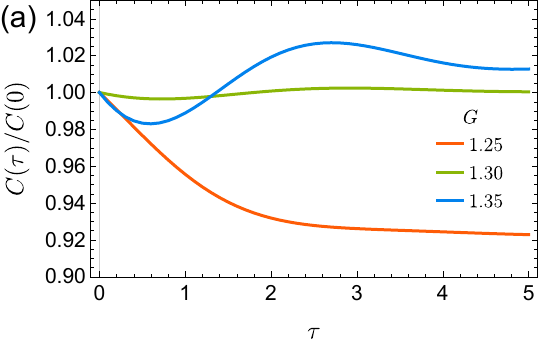}
    \includegraphics[width=0.8\linewidth]{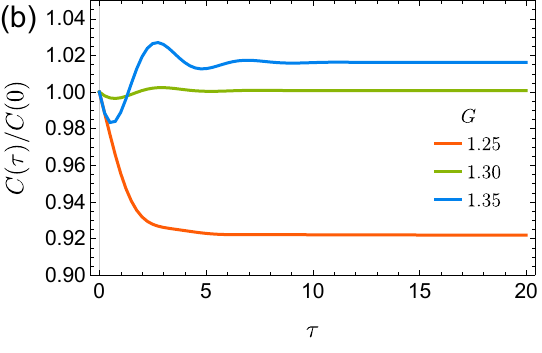}
    \caption{{\bf (a)} Short-time and {\bf (b)} long-time behaviors of the normalized correlation function $C(\tau)/C(0)$ in the Kerr model with $\Delta=1,U=1/40$ for several values of $G$. The first-order phase transition point is located at $G\approx 1.33$.}
    \label{fig:Fplot1st}
\end{figure}

We now compute the correlation function $C(\tau)$ of the current fluctuations in the photodetection measurement around the first order phase transition point ($G\approx 1.33$). Fig.~\ref{fig:Fplot1st} shows the short-time and long-time behaviors of $C(\tau)/C(0)$ for several values of $G$.
As shown in fig.~\ref{fig:Fplot1st}, we confirm that the correlation function starts oscillating along with the first order phase transition. This result indicates that the transition between an overdamped and underdamped oscillating behavior becomes discontinuous at the first order phase transitions. Note that the slow relaxation of the correlation function around first order critical points, as observed in the panel (b) of Fig.~\ref{fig:Fplot1st}, is due to the exponential divergence of the diffusion coefficient $D=S(0)$, as already discussed in \cite{kewming2022diverging}. In conclusion, we ascertain the validity of our proposal also for first-order dissipative phase transitions.

\section*{Outlook} \label{sec:conclusion}
In this work, we proposed a new approach to probe and characterize dissipative phase transitions in open quantum systems based on the correlation of output current fluctuations, its power spectrum, and the characteristic timescale of output current fluctuations. Our approach is tested and verified in the dissipative XYZ model and the driven-dissipative Kerr model. 

We first focused on the correlation function of the output current fluctuations, demonstrating a neat dynamical crossover in its behavior from an overdamped to an underdamped oscillatory behavior, where the oscillations are interpreted as signatures of the periodic transitions between two states in the quantum trajectory associated with the jump operator. This dynamical crossover is closely related to the dissipative phase transition since both stem from the competition between the coherent and incoherent dynamics. The transition in dynamics associated with the critical point is reminiscent of critical dynamics in classical systems \cite{RevModPhys.49.435}, such as ferromagnets and superfluids, although the underlying mechanisms and origins differ from those of DPTs.

By tracking the location of this dynamical crossover and performing a finite-size-scaling analysis, we were able to predict the critical point in the XYZ model. Our predictions are in excellent agreement with the mean-field result and previous estimates. In the driven-dissipative Kerr model, on the other hand, locating the dynamical crossover using this method becomes less accurate. We ascribed this deficiency to the presence of additional overdamped decay modes in the spectrum that make the prediction of the underdamped-to-overdamped transition, and hence our estimate of the critical point, inaccurate. We showed how those additional decay modes affect the accuracy and we proposed an \textit{ad hoc} solution to this problem, which is nevertheless not universally applicable since it depends on the concrete pole structure of the system.

On the other hand, we also considered the characteristic timescale of the system, defined by the time integral of the normalized correlation function of the output current fluctuations, Eq.~\eqref{diff}. This quantity displays a universal peak as a function of the system parameters, indicating the critical slowing down associated with the DPT. We proposed that, in the thermodynamic limit, the location of this peak provides a good estimate of the critical point.  
We showed that our estimate of the critical point using this second method gives excellent results in both models.

Although in this work we focused on dissipative phase transitions associated with discrete $Z_{2}$ symmetry, we expect our method based on current fluctuations to be general and to be able to capture the critical dynamics of other quantum open systems involving different symmetry breaking patterns. One key advantage of our approach is that it does not require tracking the time evolution from a specific initial condition; instead, it relies on detecting constant quantum jumps and their correlations in the steady state. 

An even more fundamental advantage of our approach relies on the fact that the dynamical properties discussed here are experimentally accessible, as demonstrated by direct measurement in optical systems \cite{fink2018signatures,beaulieu2025observation}. Interestingly, in \cite{PhysRevB.110.104303}, a classification of DPTs based on the oscillating-mode gap was recently proposed. Investigating the relation between their method and our approach based on the current fluctuations represents an important task for the near future.

Finally, in \cite{beaulieu2025observation},  first- and second-order DPTs were thoroughly characterized in an experiment on a superconducting circuit device, utilizing a criterion similar to the one proposed in this paper but based on heterodyne detection. In Appendix \ref{app:B}, we provide some preliminary results in the dissipative XYZ model using their protocol and compare to our method. A more in depth analysis of the two approaches is left for future research.

\section*{Acknowledgements}
The authors acknowledge support from the Shanghai Municipal Science and Technology Major Project (Grant No.~2019SHZDZX01). M.B. acknowledges sponsorship via the Yangyang Development Fund. 
M.M. was supported by the Shanghai Post-doctoral Excellence Program (Grant No.~2023338).

\bibliography{main}

\begin{thebibliography}{36}%
\makeatletter
\providecommand \@ifxundefined [1]{%
 \@ifx{#1\undefined}
}%
\providecommand \@ifnum [1]{%
 \ifnum #1\expandafter \@firstoftwo
 \else \expandafter \@secondoftwo
 \fi
}%
\providecommand \@ifx [1]{%
 \ifx #1\expandafter \@firstoftwo
 \else \expandafter \@secondoftwo
 \fi
}%
\providecommand \natexlab [1]{#1}%
\providecommand \enquote  [1]{``#1''}%
\providecommand \bibnamefont  [1]{#1}%
\providecommand \bibfnamefont [1]{#1}%
\providecommand \citenamefont [1]{#1}%
\providecommand \href@noop [0]{\@secondoftwo}%
\providecommand \href [0]{\begingroup \@sanitize@url \@href}%
\providecommand \@href[1]{\@@startlink{#1}\@@href}%
\providecommand \@@href[1]{\endgroup#1\@@endlink}%
\providecommand \@sanitize@url [0]{\catcode `\\12\catcode `\$12\catcode
  `\&12\catcode `\#12\catcode `\^12\catcode `\_12\catcode `\%12\relax}%
\providecommand \@@startlink[1]{}%
\providecommand \@@endlink[0]{}%
\providecommand \url  [0]{\begingroup\@sanitize@url \@url }%
\providecommand \@url [1]{\endgroup\@href {#1}{\urlprefix }}%
\providecommand \urlprefix  [0]{URL }%
\providecommand \Eprint [0]{\href }%
\providecommand \doibase [0]{https://doi.org/}%
\providecommand \selectlanguage [0]{\@gobble}%
\providecommand \bibinfo  [0]{\@secondoftwo}%
\providecommand \bibfield  [0]{\@secondoftwo}%
\providecommand \translation [1]{[#1]}%
\providecommand \BibitemOpen [0]{}%
\providecommand \bibitemStop [0]{}%
\providecommand \bibitemNoStop [0]{.\EOS\space}%
\providecommand \EOS [0]{\spacefactor3000\relax}%
\providecommand \BibitemShut  [1]{\csname bibitem#1\endcsname}%
\let\auto@bib@innerbib\@empty
\bibitem [{\citenamefont {Rivas}\ and\ \citenamefont
  {Huelga}(2012)}]{rivas2012open}%
  \BibitemOpen
  \bibfield  {author} {\bibinfo {author} {\bibfnamefont {A.}~\bibnamefont
  {Rivas}}\ and\ \bibinfo {author} {\bibfnamefont {S.~F.}\ \bibnamefont
  {Huelga}},\ }\href@noop {} {\emph {\bibinfo {title} {Open quantum
  systems}}},\ Vol.~\bibinfo {volume} {10}\ (\bibinfo  {publisher} {Springer},\
  \bibinfo {year} {2012})\BibitemShut {NoStop}%
\bibitem [{\citenamefont {Breuer}\ and\ \citenamefont
  {Petruccione}(2002)}]{breuer2002theory}%
  \BibitemOpen
  \bibfield  {author} {\bibinfo {author} {\bibfnamefont {H.-P.}\ \bibnamefont
  {Breuer}}\ and\ \bibinfo {author} {\bibfnamefont {F.}~\bibnamefont
  {Petruccione}},\ }\href@noop {} {\emph {\bibinfo {title} {The theory of open
  quantum systems}}}\ (\bibinfo  {publisher} {Oxford University Press, USA},\
  \bibinfo {year} {2002})\BibitemShut {NoStop}%
\bibitem [{\citenamefont {Rotter}\ and\ \citenamefont
  {Bird}(2015)}]{rotter2015review}%
  \BibitemOpen
  \bibfield  {author} {\bibinfo {author} {\bibfnamefont {I.}~\bibnamefont
  {Rotter}}\ and\ \bibinfo {author} {\bibfnamefont {J.}~\bibnamefont {Bird}},\
  }\bibfield  {title} {\bibinfo {title} {A review of progress in the physics of
  open quantum systems: theory and experiment},\ }\href@noop {} {\bibfield
  {journal} {\bibinfo  {journal} {Reports on Progress in Physics}\ }\textbf
  {\bibinfo {volume} {78}},\ \bibinfo {pages} {114001} (\bibinfo {year}
  {2015})}\BibitemShut {NoStop}%
\bibitem [{\citenamefont {Schaller}(2014)}]{schaller2014open}%
  \BibitemOpen
  \bibfield  {author} {\bibinfo {author} {\bibfnamefont {G.}~\bibnamefont
  {Schaller}},\ }\href@noop {} {\emph {\bibinfo {title} {Open quantum systems
  far from equilibrium}}},\ Vol.\ \bibinfo {volume} {881}\ (\bibinfo
  {publisher} {Springer},\ \bibinfo {year} {2014})\BibitemShut {NoStop}%
\bibitem [{\citenamefont {Diehl}\ \emph {et~al.}(2010)\citenamefont {Diehl},
  \citenamefont {Tomadin}, \citenamefont {Micheli}, \citenamefont {Fazio},\
  and\ \citenamefont {Zoller}}]{PhysRevLett.105.015702}%
  \BibitemOpen
  \bibfield  {author} {\bibinfo {author} {\bibfnamefont {S.}~\bibnamefont
  {Diehl}}, \bibinfo {author} {\bibfnamefont {A.}~\bibnamefont {Tomadin}},
  \bibinfo {author} {\bibfnamefont {A.}~\bibnamefont {Micheli}}, \bibinfo
  {author} {\bibfnamefont {R.}~\bibnamefont {Fazio}},\ and\ \bibinfo {author}
  {\bibfnamefont {P.}~\bibnamefont {Zoller}},\ }\bibfield  {title} {\bibinfo
  {title} {Dynamical phase transitions and instabilities in open atomic
  many-body systems},\ }\href {https://doi.org/10.1103/PhysRevLett.105.015702}
  {\bibfield  {journal} {\bibinfo  {journal} {Phys. Rev. Lett.}\ }\textbf
  {\bibinfo {volume} {105}},\ \bibinfo {pages} {015702} (\bibinfo {year}
  {2010})}\BibitemShut {NoStop}%
\bibitem [{\citenamefont {Rodriguez}\ \emph {et~al.}(2017)\citenamefont
  {Rodriguez}, \citenamefont {Casteels}, \citenamefont {Storme}, \citenamefont
  {Carlon~Zambon}, \citenamefont {Sagnes}, \citenamefont {Le~Gratiet},
  \citenamefont {Galopin}, \citenamefont {Lema\^{\i}tre}, \citenamefont {Amo},
  \citenamefont {Ciuti},\ and\ \citenamefont {Bloch}}]{PhysRevLett.118.247402}%
  \BibitemOpen
  \bibfield  {author} {\bibinfo {author} {\bibfnamefont {S.~R.~K.}\
  \bibnamefont {Rodriguez}}, \bibinfo {author} {\bibfnamefont {W.}~\bibnamefont
  {Casteels}}, \bibinfo {author} {\bibfnamefont {F.}~\bibnamefont {Storme}},
  \bibinfo {author} {\bibfnamefont {N.}~\bibnamefont {Carlon~Zambon}}, \bibinfo
  {author} {\bibfnamefont {I.}~\bibnamefont {Sagnes}}, \bibinfo {author}
  {\bibfnamefont {L.}~\bibnamefont {Le~Gratiet}}, \bibinfo {author}
  {\bibfnamefont {E.}~\bibnamefont {Galopin}}, \bibinfo {author} {\bibfnamefont
  {A.}~\bibnamefont {Lema\^{\i}tre}}, \bibinfo {author} {\bibfnamefont
  {A.}~\bibnamefont {Amo}}, \bibinfo {author} {\bibfnamefont {C.}~\bibnamefont
  {Ciuti}},\ and\ \bibinfo {author} {\bibfnamefont {J.}~\bibnamefont {Bloch}},\
  }\bibfield  {title} {\bibinfo {title} {Probing a dissipative phase transition
  via dynamical optical hysteresis},\ }\href
  {https://doi.org/10.1103/PhysRevLett.118.247402} {\bibfield  {journal}
  {\bibinfo  {journal} {Phys. Rev. Lett.}\ }\textbf {\bibinfo {volume} {118}},\
  \bibinfo {pages} {247402} (\bibinfo {year} {2017})}\BibitemShut {NoStop}%
\bibitem [{\citenamefont {Fitzpatrick}\ \emph {et~al.}(2017)\citenamefont
  {Fitzpatrick}, \citenamefont {Sundaresan}, \citenamefont {Li}, \citenamefont
  {Koch},\ and\ \citenamefont {Houck}}]{Fitzpatrick2017}%
  \BibitemOpen
  \bibfield  {author} {\bibinfo {author} {\bibfnamefont {M.}~\bibnamefont
  {Fitzpatrick}}, \bibinfo {author} {\bibfnamefont {N.~M.}\ \bibnamefont
  {Sundaresan}}, \bibinfo {author} {\bibfnamefont {A.~C.~Y.}\ \bibnamefont
  {Li}}, \bibinfo {author} {\bibfnamefont {J.}~\bibnamefont {Koch}},\ and\
  \bibinfo {author} {\bibfnamefont {A.~A.}\ \bibnamefont {Houck}},\ }\bibfield
  {title} {\bibinfo {title} {Observation of a dissipative phase transition in a
  one-dimensional circuit qed lattice},\ }\href
  {https://doi.org/10.1103/PhysRevX.7.011016} {\bibfield  {journal} {\bibinfo
  {journal} {Phys. Rev. X}\ }\textbf {\bibinfo {volume} {7}},\ \bibinfo {pages}
  {011016} (\bibinfo {year} {2017})}\BibitemShut {NoStop}%
\bibitem [{\citenamefont {Fink}\ \emph {et~al.}(2017)\citenamefont {Fink},
  \citenamefont {Dombi}, \citenamefont {Vukics}, \citenamefont {Wallraff},\
  and\ \citenamefont {Domokos}}]{Fink2017prx}%
  \BibitemOpen
  \bibfield  {author} {\bibinfo {author} {\bibfnamefont {J.~M.}\ \bibnamefont
  {Fink}}, \bibinfo {author} {\bibfnamefont {A.}~\bibnamefont {Dombi}},
  \bibinfo {author} {\bibfnamefont {A.}~\bibnamefont {Vukics}}, \bibinfo
  {author} {\bibfnamefont {A.}~\bibnamefont {Wallraff}},\ and\ \bibinfo
  {author} {\bibfnamefont {P.}~\bibnamefont {Domokos}},\ }\bibfield  {title}
  {\bibinfo {title} {Observation of the photon-blockade breakdown phase
  transition},\ }\href {https://doi.org/10.1103/PhysRevX.7.011012} {\bibfield
  {journal} {\bibinfo  {journal} {Phys. Rev. X}\ }\textbf {\bibinfo {volume}
  {7}},\ \bibinfo {pages} {011012} (\bibinfo {year} {2017})}\BibitemShut
  {NoStop}%
\bibitem [{\citenamefont {Carmichael}(2015{\natexlab{a}})}]{PhysRevX.5.031028}%
  \BibitemOpen
  \bibfield  {author} {\bibinfo {author} {\bibfnamefont {H.~J.}\ \bibnamefont
  {Carmichael}},\ }\bibfield  {title} {\bibinfo {title} {Breakdown of photon
  blockade: A dissipative quantum phase transition in zero dimensions},\ }\href
  {https://doi.org/10.1103/PhysRevX.5.031028} {\bibfield  {journal} {\bibinfo
  {journal} {Phys. Rev. X}\ }\textbf {\bibinfo {volume} {5}},\ \bibinfo {pages}
  {031028} (\bibinfo {year} {2015}{\natexlab{a}})}\BibitemShut {NoStop}%
\bibitem [{\citenamefont {Kessler}\ \emph {et~al.}(2012)\citenamefont
  {Kessler}, \citenamefont {Giedke}, \citenamefont {Imamoglu}, \citenamefont
  {Yelin}, \citenamefont {Lukin},\ and\ \citenamefont
  {Cirac}}]{Kessler:2012wsp}%
  \BibitemOpen
  \bibfield  {author} {\bibinfo {author} {\bibfnamefont {E.~M.}\ \bibnamefont
  {Kessler}}, \bibinfo {author} {\bibfnamefont {G.}~\bibnamefont {Giedke}},
  \bibinfo {author} {\bibfnamefont {A.}~\bibnamefont {Imamoglu}}, \bibinfo
  {author} {\bibfnamefont {S.~F.}\ \bibnamefont {Yelin}}, \bibinfo {author}
  {\bibfnamefont {M.~D.}\ \bibnamefont {Lukin}},\ and\ \bibinfo {author}
  {\bibfnamefont {J.~I.}\ \bibnamefont {Cirac}},\ }\bibfield  {title} {\bibinfo
  {title} {{Dissipative phase transition in a central spin system}},\ }\href
  {https://doi.org/10.1103/PhysRevA.86.012116} {\bibfield  {journal} {\bibinfo
  {journal} {Phys. Rev. A}\ }\textbf {\bibinfo {volume} {86}},\ \bibinfo
  {pages} {012116} (\bibinfo {year} {2012})}\BibitemShut {NoStop}%
\bibitem [{\citenamefont {Prosen}\ and\ \citenamefont
  {Pi\ifmmode~\check{z}\else \v{z}\fi{}orn}(2008)}]{PhysRevLett.101.105701}%
  \BibitemOpen
  \bibfield  {author} {\bibinfo {author} {\bibfnamefont {T.~c.~v.}\
  \bibnamefont {Prosen}}\ and\ \bibinfo {author} {\bibfnamefont
  {I.}~\bibnamefont {Pi\ifmmode~\check{z}\else \v{z}\fi{}orn}},\ }\bibfield
  {title} {\bibinfo {title} {Quantum phase transition in a far-from-equilibrium
  steady state of an $xy$ spin chain},\ }\href
  {https://doi.org/10.1103/PhysRevLett.101.105701} {\bibfield  {journal}
  {\bibinfo  {journal} {Phys. Rev. Lett.}\ }\textbf {\bibinfo {volume} {101}},\
  \bibinfo {pages} {105701} (\bibinfo {year} {2008})}\BibitemShut {NoStop}%
\bibitem [{\citenamefont {Mitra}\ \emph {et~al.}(2006)\citenamefont {Mitra},
  \citenamefont {Takei}, \citenamefont {Kim},\ and\ \citenamefont
  {Millis}}]{PhysRevLett.97.236808}%
  \BibitemOpen
  \bibfield  {author} {\bibinfo {author} {\bibfnamefont {A.}~\bibnamefont
  {Mitra}}, \bibinfo {author} {\bibfnamefont {S.}~\bibnamefont {Takei}},
  \bibinfo {author} {\bibfnamefont {Y.~B.}\ \bibnamefont {Kim}},\ and\ \bibinfo
  {author} {\bibfnamefont {A.~J.}\ \bibnamefont {Millis}},\ }\bibfield  {title}
  {\bibinfo {title} {Nonequilibrium quantum criticality in open electronic
  systems},\ }\href {https://doi.org/10.1103/PhysRevLett.97.236808} {\bibfield
  {journal} {\bibinfo  {journal} {Phys. Rev. Lett.}\ }\textbf {\bibinfo
  {volume} {97}},\ \bibinfo {pages} {236808} (\bibinfo {year}
  {2006})}\BibitemShut {NoStop}%
\bibitem [{\citenamefont {Minganti}\ \emph {et~al.}(2018)\citenamefont
  {Minganti}, \citenamefont {Biella}, \citenamefont {Bartolo},\ and\
  \citenamefont {Ciuti}}]{Minganti:2018kgs}%
  \BibitemOpen
  \bibfield  {author} {\bibinfo {author} {\bibfnamefont {F.}~\bibnamefont
  {Minganti}}, \bibinfo {author} {\bibfnamefont {A.}~\bibnamefont {Biella}},
  \bibinfo {author} {\bibfnamefont {N.}~\bibnamefont {Bartolo}},\ and\ \bibinfo
  {author} {\bibfnamefont {C.}~\bibnamefont {Ciuti}},\ }\bibfield  {title}
  {\bibinfo {title} {{Spectral theory of Liouvillians for dissipative phase
  transitions}},\ }\href {https://doi.org/10.1103/PhysRevA.98.042118}
  {\bibfield  {journal} {\bibinfo  {journal} {Phys. Rev. A}\ }\textbf {\bibinfo
  {volume} {98}},\ \bibinfo {pages} {042118} (\bibinfo {year}
  {2018})}\BibitemShut {NoStop}%
\bibitem [{\citenamefont {Carmichael}(2015{\natexlab{b}})}]{Charmichael2015}%
  \BibitemOpen
  \bibfield  {author} {\bibinfo {author} {\bibfnamefont {H.~J.}\ \bibnamefont
  {Carmichael}},\ }\bibfield  {title} {\bibinfo {title} {Breakdown of photon
  blockade: A dissipative quantum phase transition in zero dimensions},\ }\href
  {https://doi.org/10.1103/PhysRevX.5.031028} {\bibfield  {journal} {\bibinfo
  {journal} {Phys. Rev. X}\ }\textbf {\bibinfo {volume} {5}},\ \bibinfo {pages}
  {031028} (\bibinfo {year} {2015}{\natexlab{b}})}\BibitemShut {NoStop}%
\bibitem [{\citenamefont {Bartolo}\ \emph {et~al.}(2016)\citenamefont
  {Bartolo}, \citenamefont {Minganti}, \citenamefont {Casteels},\ and\
  \citenamefont {Ciuti}}]{Bartolo2016}%
  \BibitemOpen
  \bibfield  {author} {\bibinfo {author} {\bibfnamefont {N.}~\bibnamefont
  {Bartolo}}, \bibinfo {author} {\bibfnamefont {F.}~\bibnamefont {Minganti}},
  \bibinfo {author} {\bibfnamefont {W.}~\bibnamefont {Casteels}},\ and\
  \bibinfo {author} {\bibfnamefont {C.}~\bibnamefont {Ciuti}},\ }\bibfield
  {title} {\bibinfo {title} {Exact steady state of a kerr resonator with one-
  and two-photon driving and dissipation: Controllable wigner-function
  multimodality and dissipative phase transitions},\ }\href
  {https://doi.org/10.1103/PhysRevA.94.033841} {\bibfield  {journal} {\bibinfo
  {journal} {Phys. Rev. A}\ }\textbf {\bibinfo {volume} {94}},\ \bibinfo
  {pages} {033841} (\bibinfo {year} {2016})}\BibitemShut {NoStop}%
\bibitem [{\citenamefont {Mendoza-Arenas}\ \emph {et~al.}(2016)\citenamefont
  {Mendoza-Arenas}, \citenamefont {Clark}, \citenamefont {Felicetti},
  \citenamefont {Romero}, \citenamefont {Solano}, \citenamefont {Angelakis},\
  and\ \citenamefont {Jaksch}}]{Mendoza2016}%
  \BibitemOpen
  \bibfield  {author} {\bibinfo {author} {\bibfnamefont {J.~J.}\ \bibnamefont
  {Mendoza-Arenas}}, \bibinfo {author} {\bibfnamefont {S.~R.}\ \bibnamefont
  {Clark}}, \bibinfo {author} {\bibfnamefont {S.}~\bibnamefont {Felicetti}},
  \bibinfo {author} {\bibfnamefont {G.}~\bibnamefont {Romero}}, \bibinfo
  {author} {\bibfnamefont {E.}~\bibnamefont {Solano}}, \bibinfo {author}
  {\bibfnamefont {D.~G.}\ \bibnamefont {Angelakis}},\ and\ \bibinfo {author}
  {\bibfnamefont {D.}~\bibnamefont {Jaksch}},\ }\bibfield  {title} {\bibinfo
  {title} {Beyond mean-field bistability in driven-dissipative lattices:
  Bunching-antibunching transition and quantum simulation},\ }\href
  {https://doi.org/10.1103/PhysRevA.93.023821} {\bibfield  {journal} {\bibinfo
  {journal} {Phys. Rev. A}\ }\textbf {\bibinfo {volume} {93}},\ \bibinfo
  {pages} {023821} (\bibinfo {year} {2016})}\BibitemShut {NoStop}%
\bibitem [{\citenamefont {Biella}\ \emph {et~al.}(2017)\citenamefont {Biella},
  \citenamefont {Storme}, \citenamefont {Lebreuilly}, \citenamefont {Rossini},
  \citenamefont {Fazio}, \citenamefont {Carusotto},\ and\ \citenamefont
  {Ciuti}}]{Biella2017}%
  \BibitemOpen
  \bibfield  {author} {\bibinfo {author} {\bibfnamefont {A.}~\bibnamefont
  {Biella}}, \bibinfo {author} {\bibfnamefont {F.}~\bibnamefont {Storme}},
  \bibinfo {author} {\bibfnamefont {J.}~\bibnamefont {Lebreuilly}}, \bibinfo
  {author} {\bibfnamefont {D.}~\bibnamefont {Rossini}}, \bibinfo {author}
  {\bibfnamefont {R.}~\bibnamefont {Fazio}}, \bibinfo {author} {\bibfnamefont
  {I.}~\bibnamefont {Carusotto}},\ and\ \bibinfo {author} {\bibfnamefont
  {C.}~\bibnamefont {Ciuti}},\ }\bibfield  {title} {\bibinfo {title} {Phase
  diagram of incoherently driven strongly correlated photonic lattices},\
  }\href {https://doi.org/10.1103/PhysRevA.96.023839} {\bibfield  {journal}
  {\bibinfo  {journal} {Phys. Rev. A}\ }\textbf {\bibinfo {volume} {96}},\
  \bibinfo {pages} {023839} (\bibinfo {year} {2017})}\BibitemShut {NoStop}%
\bibitem [{\citenamefont {Savona}(2017)}]{Savona2017}%
  \BibitemOpen
  \bibfield  {author} {\bibinfo {author} {\bibfnamefont {V.}~\bibnamefont
  {Savona}},\ }\bibfield  {title} {\bibinfo {title} {Spontaneous symmetry
  breaking in a quadratically driven nonlinear photonic lattice},\ }\href
  {https://doi.org/10.1103/PhysRevA.96.033826} {\bibfield  {journal} {\bibinfo
  {journal} {Phys. Rev. A}\ }\textbf {\bibinfo {volume} {96}},\ \bibinfo
  {pages} {033826} (\bibinfo {year} {2017})}\BibitemShut {NoStop}%
\bibitem [{\citenamefont {Foss-Feig}\ \emph {et~al.}(2017)\citenamefont
  {Foss-Feig}, \citenamefont {Niroula}, \citenamefont {Young}, \citenamefont
  {Hafezi}, \citenamefont {Gorshkov}, \citenamefont {Wilson},\ and\
  \citenamefont {Maghrebi}}]{FossFeig2017}%
  \BibitemOpen
  \bibfield  {author} {\bibinfo {author} {\bibfnamefont {M.}~\bibnamefont
  {Foss-Feig}}, \bibinfo {author} {\bibfnamefont {P.}~\bibnamefont {Niroula}},
  \bibinfo {author} {\bibfnamefont {J.~T.}\ \bibnamefont {Young}}, \bibinfo
  {author} {\bibfnamefont {M.}~\bibnamefont {Hafezi}}, \bibinfo {author}
  {\bibfnamefont {A.~V.}\ \bibnamefont {Gorshkov}}, \bibinfo {author}
  {\bibfnamefont {R.~M.}\ \bibnamefont {Wilson}},\ and\ \bibinfo {author}
  {\bibfnamefont {M.~F.}\ \bibnamefont {Maghrebi}},\ }\bibfield  {title}
  {\bibinfo {title} {Emergent equilibrium in many-body optical bistability},\
  }\href {https://doi.org/10.1103/PhysRevA.95.043826} {\bibfield  {journal}
  {\bibinfo  {journal} {Phys. Rev. A}\ }\textbf {\bibinfo {volume} {95}},\
  \bibinfo {pages} {043826} (\bibinfo {year} {2017})}\BibitemShut {NoStop}%
\bibitem [{\citenamefont {Vicentini}\ \emph {et~al.}(2018)\citenamefont
  {Vicentini}, \citenamefont {Minganti}, \citenamefont {Rota}, \citenamefont
  {Orso},\ and\ \citenamefont {Ciuti}}]{Vicentini2018}%
  \BibitemOpen
  \bibfield  {author} {\bibinfo {author} {\bibfnamefont {F.}~\bibnamefont
  {Vicentini}}, \bibinfo {author} {\bibfnamefont {F.}~\bibnamefont {Minganti}},
  \bibinfo {author} {\bibfnamefont {R.}~\bibnamefont {Rota}}, \bibinfo {author}
  {\bibfnamefont {G.}~\bibnamefont {Orso}},\ and\ \bibinfo {author}
  {\bibfnamefont {C.}~\bibnamefont {Ciuti}},\ }\bibfield  {title} {\bibinfo
  {title} {Critical slowing down in driven-dissipative bose-hubbard lattices},\
  }\href {https://doi.org/10.1103/PhysRevA.97.013853} {\bibfield  {journal}
  {\bibinfo  {journal} {Phys. Rev. A}\ }\textbf {\bibinfo {volume} {97}},\
  \bibinfo {pages} {013853} (\bibinfo {year} {2018})}\BibitemShut {NoStop}%
\bibitem [{\citenamefont {Lee}\ \emph {et~al.}(2013)\citenamefont {Lee},
  \citenamefont {Gopalakrishnan},\ and\ \citenamefont {Lukin}}]{Lee2013}%
  \BibitemOpen
  \bibfield  {author} {\bibinfo {author} {\bibfnamefont {T.~E.}\ \bibnamefont
  {Lee}}, \bibinfo {author} {\bibfnamefont {S.}~\bibnamefont
  {Gopalakrishnan}},\ and\ \bibinfo {author} {\bibfnamefont {M.~D.}\
  \bibnamefont {Lukin}},\ }\bibfield  {title} {\bibinfo {title} {Unconventional
  magnetism via optical pumping of interacting spin systems},\ }\href
  {https://doi.org/10.1103/PhysRevLett.110.257204} {\bibfield  {journal}
  {\bibinfo  {journal} {Phys. Rev. Lett.}\ }\textbf {\bibinfo {volume} {110}},\
  \bibinfo {pages} {257204} (\bibinfo {year} {2013})}\BibitemShut {NoStop}%
\bibitem [{\citenamefont {Jin}\ \emph {et~al.}(2016)\citenamefont {Jin},
  \citenamefont {Biella}, \citenamefont {Viyuela}, \citenamefont {Mazza},
  \citenamefont {Keeling}, \citenamefont {Fazio},\ and\ \citenamefont
  {Rossini}}]{Jin2016}%
  \BibitemOpen
  \bibfield  {author} {\bibinfo {author} {\bibfnamefont {J.}~\bibnamefont
  {Jin}}, \bibinfo {author} {\bibfnamefont {A.}~\bibnamefont {Biella}},
  \bibinfo {author} {\bibfnamefont {O.}~\bibnamefont {Viyuela}}, \bibinfo
  {author} {\bibfnamefont {L.}~\bibnamefont {Mazza}}, \bibinfo {author}
  {\bibfnamefont {J.}~\bibnamefont {Keeling}}, \bibinfo {author} {\bibfnamefont
  {R.}~\bibnamefont {Fazio}},\ and\ \bibinfo {author} {\bibfnamefont
  {D.}~\bibnamefont {Rossini}},\ }\bibfield  {title} {\bibinfo {title} {Cluster
  mean-field approach to the steady-state phase diagram of dissipative spin
  systems},\ }\href {https://doi.org/10.1103/PhysRevX.6.031011} {\bibfield
  {journal} {\bibinfo  {journal} {Phys. Rev. X}\ }\textbf {\bibinfo {volume}
  {6}},\ \bibinfo {pages} {031011} (\bibinfo {year} {2016})}\BibitemShut
  {NoStop}%
\bibitem [{\citenamefont {Rota}\ \emph {et~al.}(2017)\citenamefont {Rota},
  \citenamefont {Storme}, \citenamefont {Bartolo}, \citenamefont {Fazio},\ and\
  \citenamefont {Ciuti}}]{Rota2017}%
  \BibitemOpen
  \bibfield  {author} {\bibinfo {author} {\bibfnamefont {R.}~\bibnamefont
  {Rota}}, \bibinfo {author} {\bibfnamefont {F.}~\bibnamefont {Storme}},
  \bibinfo {author} {\bibfnamefont {N.}~\bibnamefont {Bartolo}}, \bibinfo
  {author} {\bibfnamefont {R.}~\bibnamefont {Fazio}},\ and\ \bibinfo {author}
  {\bibfnamefont {C.}~\bibnamefont {Ciuti}},\ }\bibfield  {title} {\bibinfo
  {title} {Critical behavior of dissipative two-dimensional spin lattices},\
  }\href {https://doi.org/10.1103/PhysRevB.95.134431} {\bibfield  {journal}
  {\bibinfo  {journal} {Phys. Rev. B}\ }\textbf {\bibinfo {volume} {95}},\
  \bibinfo {pages} {134431} (\bibinfo {year} {2017})}\BibitemShut {NoStop}%
\bibitem [{\citenamefont {Rota}\ \emph {et~al.}(2018)\citenamefont {Rota},
  \citenamefont {Minganti}, \citenamefont {Biella},\ and\ \citenamefont
  {Ciuti}}]{rota2018dynamical}%
  \BibitemOpen
  \bibfield  {author} {\bibinfo {author} {\bibfnamefont {R.}~\bibnamefont
  {Rota}}, \bibinfo {author} {\bibfnamefont {F.}~\bibnamefont {Minganti}},
  \bibinfo {author} {\bibfnamefont {A.}~\bibnamefont {Biella}},\ and\ \bibinfo
  {author} {\bibfnamefont {C.}~\bibnamefont {Ciuti}},\ }\bibfield  {title}
  {\bibinfo {title} {Dynamical properties of dissipative xyz heisenberg
  lattices},\ }\href {https://doi.org/10.1088/1367-2630/aab703} {\bibfield
  {journal} {\bibinfo  {journal} {New Journal of Physics}\ }\textbf {\bibinfo
  {volume} {20}},\ \bibinfo {pages} {045003} (\bibinfo {year}
  {2018})}\BibitemShut {NoStop}%
\bibitem [{\citenamefont {Li}\ \emph {et~al.}(2022)\citenamefont {Li},
  \citenamefont {Li},\ and\ \citenamefont {Jin}}]{Li2022}%
  \BibitemOpen
  \bibfield  {author} {\bibinfo {author} {\bibfnamefont {X.}~\bibnamefont
  {Li}}, \bibinfo {author} {\bibfnamefont {Y.}~\bibnamefont {Li}},\ and\
  \bibinfo {author} {\bibfnamefont {J.}~\bibnamefont {Jin}},\ }\bibfield
  {title} {\bibinfo {title} {Steady-state susceptibility in continuous phase
  transitions of dissipative systems},\ }\href
  {https://doi.org/10.1103/PhysRevA.105.052226} {\bibfield  {journal} {\bibinfo
   {journal} {Phys. Rev. A}\ }\textbf {\bibinfo {volume} {105}},\ \bibinfo
  {pages} {052226} (\bibinfo {year} {2022})}\BibitemShut {NoStop}%
\bibitem [{\citenamefont {Haga}(2024)}]{PhysRevB.110.104303}%
  \BibitemOpen
  \bibfield  {author} {\bibinfo {author} {\bibfnamefont {T.}~\bibnamefont
  {Haga}},\ }\bibfield  {title} {\bibinfo {title} {Oscillating-mode gap: An
  indicator of phase transitions in open quantum many-body systems},\ }\href
  {https://doi.org/10.1103/PhysRevB.110.104303} {\bibfield  {journal} {\bibinfo
   {journal} {Phys. Rev. B}\ }\textbf {\bibinfo {volume} {110}},\ \bibinfo
  {pages} {104303} (\bibinfo {year} {2024})}\BibitemShut {NoStop}%
\bibitem [{\citenamefont {Landi}\ \emph {et~al.}(2024)\citenamefont {Landi},
  \citenamefont {Kewming}, \citenamefont {Mitchison},\ and\ \citenamefont
  {Potts}}]{Landi:2023ktg}%
  \BibitemOpen
  \bibfield  {author} {\bibinfo {author} {\bibfnamefont {G.~T.}\ \bibnamefont
  {Landi}}, \bibinfo {author} {\bibfnamefont {M.~J.}\ \bibnamefont {Kewming}},
  \bibinfo {author} {\bibfnamefont {M.~T.}\ \bibnamefont {Mitchison}},\ and\
  \bibinfo {author} {\bibfnamefont {P.~P.}\ \bibnamefont {Potts}},\ }\bibfield
  {title} {\bibinfo {title} {{Current Fluctuations in Open Quantum Systems:
  Bridging the Gap Between Quantum Continuous Measurements and Full Counting
  Statistics}},\ }\href {https://doi.org/10.1103/PRXQuantum.5.020201}
  {\bibfield  {journal} {\bibinfo  {journal} {PRX Quantum}\ }\textbf {\bibinfo
  {volume} {5}},\ \bibinfo {pages} {020201} (\bibinfo {year} {2024})},\ \Eprint
  {https://arxiv.org/abs/2303.04270} {arXiv:2303.04270 [quant-ph]} \BibitemShut
  {NoStop}%
\bibitem [{\citenamefont {Fink}\ \emph {et~al.}(2018)\citenamefont {Fink},
  \citenamefont {Schade}, \citenamefont {H{\"o}fling}, \citenamefont
  {Schneider},\ and\ \citenamefont {Imamoglu}}]{fink2018signatures}%
  \BibitemOpen
  \bibfield  {author} {\bibinfo {author} {\bibfnamefont {T.}~\bibnamefont
  {Fink}}, \bibinfo {author} {\bibfnamefont {A.}~\bibnamefont {Schade}},
  \bibinfo {author} {\bibfnamefont {S.}~\bibnamefont {H{\"o}fling}}, \bibinfo
  {author} {\bibfnamefont {C.}~\bibnamefont {Schneider}},\ and\ \bibinfo
  {author} {\bibfnamefont {A.}~\bibnamefont {Imamoglu}},\ }\bibfield  {title}
  {\bibinfo {title} {Signatures of a dissipative phase transition in photon
  correlation measurements},\ }\href@noop {} {\bibfield  {journal} {\bibinfo
  {journal} {Nature Physics}\ }\textbf {\bibinfo {volume} {14}},\ \bibinfo
  {pages} {365} (\bibinfo {year} {2018})}\BibitemShut {NoStop}%
\bibitem [{\citenamefont {Beaulieu}\ \emph {et~al.}(2025)\citenamefont
  {Beaulieu}, \citenamefont {Minganti}, \citenamefont {Frasca}, \citenamefont
  {Savona}, \citenamefont {Felicetti}, \citenamefont {Di~Candia},\ and\
  \citenamefont {Scarlino}}]{beaulieu2025observation}%
  \BibitemOpen
  \bibfield  {author} {\bibinfo {author} {\bibfnamefont {G.}~\bibnamefont
  {Beaulieu}}, \bibinfo {author} {\bibfnamefont {F.}~\bibnamefont {Minganti}},
  \bibinfo {author} {\bibfnamefont {S.}~\bibnamefont {Frasca}}, \bibinfo
  {author} {\bibfnamefont {V.}~\bibnamefont {Savona}}, \bibinfo {author}
  {\bibfnamefont {S.}~\bibnamefont {Felicetti}}, \bibinfo {author}
  {\bibfnamefont {R.}~\bibnamefont {Di~Candia}},\ and\ \bibinfo {author}
  {\bibfnamefont {P.}~\bibnamefont {Scarlino}},\ }\bibfield  {title} {\bibinfo
  {title} {Observation of first-and second-order dissipative phase transitions
  in a two-photon driven kerr resonator},\ }\href@noop {} {\bibfield  {journal}
  {\bibinfo  {journal} {Nature Communications}\ }\textbf {\bibinfo {volume}
  {16}},\ \bibinfo {pages} {1954} (\bibinfo {year} {2025})}\BibitemShut
  {NoStop}%
\bibitem [{\citenamefont {Campaioli}\ \emph {et~al.}(2024)\citenamefont
  {Campaioli}, \citenamefont {Cole},\ and\ \citenamefont
  {Hapuarachchi}}]{PRXQuantum.5.020202}%
  \BibitemOpen
  \bibfield  {author} {\bibinfo {author} {\bibfnamefont {F.}~\bibnamefont
  {Campaioli}}, \bibinfo {author} {\bibfnamefont {J.~H.}\ \bibnamefont
  {Cole}},\ and\ \bibinfo {author} {\bibfnamefont {H.}~\bibnamefont
  {Hapuarachchi}},\ }\bibfield  {title} {\bibinfo {title} {Quantum master
  equations: Tips and tricks for quantum optics, quantum computing, and
  beyond},\ }\href {https://doi.org/10.1103/PRXQuantum.5.020202} {\bibfield
  {journal} {\bibinfo  {journal} {PRX Quantum}\ }\textbf {\bibinfo {volume}
  {5}},\ \bibinfo {pages} {020202} (\bibinfo {year} {2024})}\BibitemShut
  {NoStop}%
\bibitem [{\citenamefont {Cai}\ and\ \citenamefont {Barthel}(2013)}]{Cai2013}%
  \BibitemOpen
  \bibfield  {author} {\bibinfo {author} {\bibfnamefont {Z.}~\bibnamefont
  {Cai}}\ and\ \bibinfo {author} {\bibfnamefont {T.}~\bibnamefont {Barthel}},\
  }\bibfield  {title} {\bibinfo {title} {Algebraic versus exponential
  decoherence in dissipative many-particle systems},\ }\href
  {https://doi.org/10.1103/PhysRevLett.111.150403} {\bibfield  {journal}
  {\bibinfo  {journal} {Phys. Rev. Lett.}\ }\textbf {\bibinfo {volume} {111}},\
  \bibinfo {pages} {150403} (\bibinfo {year} {2013})}\BibitemShut {NoStop}%
\bibitem [{\citenamefont {Ren}\ \emph {et~al.}(2020)\citenamefont {Ren},
  \citenamefont {Li}, \citenamefont {Li}, \citenamefont {Cai},\ and\
  \citenamefont {Wang}}]{Ren2020}%
  \BibitemOpen
  \bibfield  {author} {\bibinfo {author} {\bibfnamefont {J.}~\bibnamefont
  {Ren}}, \bibinfo {author} {\bibfnamefont {Q.}~\bibnamefont {Li}}, \bibinfo
  {author} {\bibfnamefont {W.}~\bibnamefont {Li}}, \bibinfo {author}
  {\bibfnamefont {Z.}~\bibnamefont {Cai}},\ and\ \bibinfo {author}
  {\bibfnamefont {X.}~\bibnamefont {Wang}},\ }\bibfield  {title} {\bibinfo
  {title} {Noise-driven universal dynamics towards an infinite temperature
  state},\ }\href {https://doi.org/10.1103/PhysRevLett.124.130602} {\bibfield
  {journal} {\bibinfo  {journal} {Phys. Rev. Lett.}\ }\textbf {\bibinfo
  {volume} {124}},\ \bibinfo {pages} {130602} (\bibinfo {year}
  {2020})}\BibitemShut {NoStop}%
\bibitem [{\citenamefont {Biella}\ \emph {et~al.}(2018)\citenamefont {Biella},
  \citenamefont {Jin}, \citenamefont {Viyuela}, \citenamefont {Ciuti},
  \citenamefont {Fazio},\ and\ \citenamefont {Rossini}}]{PhysRevB.97.035103}%
  \BibitemOpen
  \bibfield  {author} {\bibinfo {author} {\bibfnamefont {A.}~\bibnamefont
  {Biella}}, \bibinfo {author} {\bibfnamefont {J.}~\bibnamefont {Jin}},
  \bibinfo {author} {\bibfnamefont {O.}~\bibnamefont {Viyuela}}, \bibinfo
  {author} {\bibfnamefont {C.}~\bibnamefont {Ciuti}}, \bibinfo {author}
  {\bibfnamefont {R.}~\bibnamefont {Fazio}},\ and\ \bibinfo {author}
  {\bibfnamefont {D.}~\bibnamefont {Rossini}},\ }\bibfield  {title} {\bibinfo
  {title} {Linked cluster expansions for open quantum systems on a lattice},\
  }\href {https://doi.org/10.1103/PhysRevB.97.035103} {\bibfield  {journal}
  {\bibinfo  {journal} {Phys. Rev. B}\ }\textbf {\bibinfo {volume} {97}},\
  \bibinfo {pages} {035103} (\bibinfo {year} {2018})}\BibitemShut {NoStop}%
\bibitem [{\citenamefont {Kewming}\ \emph {et~al.}(2022)\citenamefont
  {Kewming}, \citenamefont {Mitchison},\ and\ \citenamefont
  {Landi}}]{kewming2022diverging}%
  \BibitemOpen
  \bibfield  {author} {\bibinfo {author} {\bibfnamefont {M.~J.}\ \bibnamefont
  {Kewming}}, \bibinfo {author} {\bibfnamefont {M.~T.}\ \bibnamefont
  {Mitchison}},\ and\ \bibinfo {author} {\bibfnamefont {G.~T.}\ \bibnamefont
  {Landi}},\ }\bibfield  {title} {\bibinfo {title} {Diverging current
  fluctuations in critical kerr resonators},\ }\href
  {https://doi.org/10.1103/PhysRevA.106.033707} {\bibfield  {journal} {\bibinfo
   {journal} {Phys. Rev. A}\ }\textbf {\bibinfo {volume} {106}},\ \bibinfo
  {pages} {033707} (\bibinfo {year} {2022})}\BibitemShut {NoStop}%
\bibitem [{\citenamefont {Zhang}\ and\ \citenamefont
  {Baranger}(2021)}]{Zhang2021}%
  \BibitemOpen
  \bibfield  {author} {\bibinfo {author} {\bibfnamefont {X.~H.~H.}\
  \bibnamefont {Zhang}}\ and\ \bibinfo {author} {\bibfnamefont {H.~U.}\
  \bibnamefont {Baranger}},\ }\bibfield  {title} {\bibinfo {title}
  {Driven-dissipative phase transition in a kerr oscillator: From semiclassical
  $\mathcal{PT}$ symmetry to quantum fluctuations},\ }\href
  {https://doi.org/10.1103/PhysRevA.103.033711} {\bibfield  {journal} {\bibinfo
   {journal} {Phys. Rev. A}\ }\textbf {\bibinfo {volume} {103}},\ \bibinfo
  {pages} {033711} (\bibinfo {year} {2021})}\BibitemShut {NoStop}%
\bibitem [{\citenamefont {Hohenberg}\ and\ \citenamefont
  {Halperin}(1977)}]{RevModPhys.49.435}%
  \BibitemOpen
  \bibfield  {author} {\bibinfo {author} {\bibfnamefont {P.~C.}\ \bibnamefont
  {Hohenberg}}\ and\ \bibinfo {author} {\bibfnamefont {B.~I.}\ \bibnamefont
  {Halperin}},\ }\bibfield  {title} {\bibinfo {title} {Theory of dynamic
  critical phenomena},\ }\href {https://doi.org/10.1103/RevModPhys.49.435}
  {\bibfield  {journal} {\bibinfo  {journal} {Rev. Mod. Phys.}\ }\textbf
  {\bibinfo {volume} {49}},\ \bibinfo {pages} {435} (\bibinfo {year}
  {1977})}\BibitemShut {NoStop}%
\end{thebibliography}%

\appendix
\section{Mean-field analysis of the dissipative XYZ model} \label{app:A}
In this appendix, we review the mean-field analysis of the dissipative XYZ model \cite{Lee2013} and show the corresponding results for the correlation function.
In the mean-field approximation, we assume that the density matrix of the total system can be factorized as 
\begin{equation}
    \hat{\rho} = \bigotimes_{j} \hat{\rho}_{j},
\end{equation}
and the total system is reduced to a single-qubit system.
As a result, the QME can be written as
\begin{equation}
    \frac{\dd \hat{\rho}}{\dd t} = - i [\hat{H}_{\rm MF}, \hat{\rho}] +\gamma \bigg[ \hat{\sigma}^{-}\hat{\rho}\hat{\sigma}^{+} - \frac{1}{2}\{ \hat{\sigma}^{+}\hat{\sigma}^{-}, \hat{\rho} \} \bigg], \label{eq:QMEMF}
\end{equation}
where $\hat{H}_{\rm MF}$ is defined by \eqref{eq:HMF}. The expectation values of the Pauli matrix satisfy
\begin{eqnarray}
    \frac{\dd}{\dd t} \expval{\hat{\sigma}^{x}} &=& 8(J_{y}-J_{z}) \expval{\hat{\sigma}^{y}}\expval{\hat{\sigma}^{z}} - \frac{\gamma}{2}\expval{\hat{\sigma}^{x}}, \\
    \frac{\dd}{\dd t}\expval{\hat{\sigma}^{y}} &=& 8(J_{z}-J_{x}) \expval{\hat{\sigma}^{x}}\expval{\hat{\sigma}^{z}} - \frac{\gamma}{2}\expval{\hat{\sigma}^{y}}, \\
    \frac{\dd}{\dd t}\expval{\hat{\sigma}^{z}} &=& 8(J_{x}-J_{y})\expval{\hat{\sigma}^{x}}\expval{\hat{\sigma}^{y}} - \gamma (\expval{\hat{\sigma}^{z}}+1),
\end{eqnarray}
where we also assume that the system exists in a two-dimensional square lattice.

Here, $\expval{\hat{\sigma}^{x,y}}=0$ and $\expval{\hat{\sigma}^{z}}=-1$ is always a solution in a steady state, corresponding to the state with all spins down.
One can also find the steady -state solution with finite $\expval{\hat{\sigma}^{x,y}}$.
The explicit form of the solution is
\begin{eqnarray}
    \expval{\hat{\sigma}^{x}} &=& \frac{J_{x}\sqrt{\gamma}}{8\sqrt{2}\sqrt{J_{z}-J_{x}}} \sqrt{\frac{16\sqrt{(J_{y}-J_{z})(J_{z}-J_{x})}-\gamma}{J_{y}-J_{x}}}, \nonumber \\ \label{eq:sxMF} \\
    \expval{\hat{\sigma}^{y}} &=& \frac{J_{y}\sqrt{\gamma}}{8\sqrt{2}\sqrt{J_{y}-J_{x}}} \sqrt{\frac{16\sqrt{(J_{y}-J_{z})(J_{z}-J_{x})}-\gamma}{J_{y}-J_{z}}}, \nonumber \\ \label{eq:syMF} \\
    \expval{\hat{\sigma}^{z}} &=& -\frac{J_{z}\gamma}{16\sqrt{(J_{y}-J_{z})(J_{z}-J_{x})}},
\end{eqnarray}
where we assume $J_{y}>J_{z}>J_{x}$, which is always satisfied for our choice of parameters.
The zero of (\ref{eq:sxMF}) or (\ref{eq:syMF}) determines the critical point value reported in the main text in Eq.~\eqref{eq:Jyc}. With $J_{x}=9/10$ and $J_{z}=1$, we obtain $J_{y}^{c} = 133/128 \approx 1.04$.

The value of $J_{y}^{\rm peak}$ is derived from Eq.~(\ref{eq:omegapeak}). 
Approximating the Rabi frequency as $\Omega_{\rm R} = J_{x}\expval{\hat{\sigma}^{x}}$ in the vicinity of the critical point with $J_{x}=9/10$ and $J_{z}=1$, the condition for the emergence of the peak in the power spectrum is given by
\begin{equation}
    \Omega_{\rm R}^{\rm MF} > \frac{1}{2} \quad \Longleftrightarrow \quad \frac{162\sqrt{2}}{5\sqrt{445}} \sqrt{J_{y} - J_{y}^{c}} >\frac{1}{2}.
\end{equation}
By direct computation, we then obtain $J_{y}^{\rm peak} = J_{y}^{c} +11125/209952 \approx 1.09$.

\begin{figure}[tbp]
    \centering
    \includegraphics[width=0.9\linewidth]{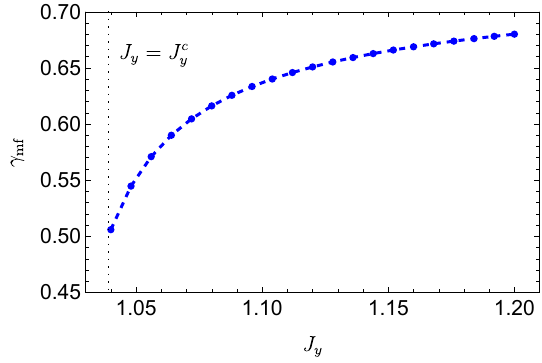}
    \caption{Relaxation rate of the oscillatory mode, $\gamma_{\rm mf}$, as a function of $J_{y}$ in the mean-field approximation. The black dotted line denotes the critical point $J_{y}=J_{y}^{c}$. The blue dashed line denotes the analytical solution from Eq. \eqref{eq:MFeig}.}
    \label{fig:MFtau}
\end{figure}

In Fig.~\ref{fig:FMF} in the main text, we have shown the oscillatory damped behavior that emerges in the correlation function $C(\tau)$ within the mean-field approximation. Here, in Fig.~\ref{fig:MFtau}, we also show the relaxation rate of the corresponding oscillatory mode, $\gamma_{\rm mf}$, as a function of $J_{y}$. The relaxation rate is obtained by using the fitting formula
\begin{equation}
    C(\tau)= c_0 \cos (\omega_{\rm mf}t) \exp(-\gamma_{\rm mf} t),
\end{equation}
where $c_0$ is a normalization constant, $\omega_{\rm mf}$ the frequency of oscillations, and $\gamma_{\rm mf}$ the relaxation rate.
The relaxation rate decreases towards the critical point, indicating that the relaxation time becomes longer, a behavior reminiscent of the critical slowing down. Nevertheless, we notice that this relaxation time does not diverge at the critical point, but attains a finite value. This is because, in the mean-field approximation, the Hamiltonian (and the Liouvillian as well) depends not only on the system parameters but also on the density matrix, as defined in \eqref{eq:HMF}. In fact, in order to obtain the Liouvillian for the steady state, we need to solve the self-consistent equation \eqref{eq:QMEMF} for the density matrix. In the process of converging to the steady state, the relaxation time is divergent at the critical point. However, the resulting Liouvillian spectrum replaced by the steady state density matrix does not show any divergent time-scale at the critical point, which is evident from the solution of Eq. \eqref{eq:MFeig} below.

Finally, in the mean-field approximation, we find that the frequency of the oscillations in $C(\tau)$, $\omega_{\rm mf}$, is exactly determined by the eigenvalues of the Liouvillian with the steady state density matrix, obtained from solving the equation
\begin{align}
    \lambda^{3} &+ 4 \gamma \lambda^{2} + 256\left(J_{x}^{2}S_{x}^{2}+J_{y}^{2}S_{y}^{2} + J_{z}^{2} S_{z}^{2} +\frac{5}{256}\gamma^{2} \right)\lambda \nonumber\\
    &+256\gamma \left(J_{x}^{2}S_{x}^{2}+J_{y}^{2}S_{y}^{2} + 2 J_{z}^{2} S_{z}^{2} + \frac{\gamma^{2}}{128} \right) =0, \label{eq:MFeig}
\end{align}
where $S_{\alpha} \equiv \expval{\hat{\sigma}^{\alpha}}$, with $\alpha =x,y,z$. The blue line in Fig.~\ref{fig:MFtau} denotes the real part of the complex eigenvalue, which determines the relaxation rate. This result indicates that the dynamical behavior of the correlation function is closely related to the Liouvillian spectrum. 

\begin{figure}
    \centering
    \includegraphics[width=0.9\linewidth]{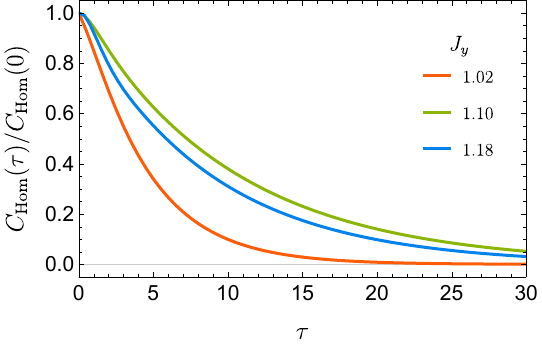}
    \caption{Normalized correlation function $C_{\rm Hom}(\tau)/C_{\rm Hom}(0)$ as a function of $\tau$ for two values of $J_{y}$. The results are for a $3\times 3$ square lattice with $J_{x}=0.9$ and $J_{z}=1$.}
    \label{fig:FplotHomo}
\end{figure}

\section{Homodyne detection} \label{app:B}
In the main text, we restricted our analysis to current fluctuations based on photodetection measurements. In this appendix, we turn to homodyne measurement, corresponding to a different unraveling of the quantum master equation. For details, we refer to Ref. \cite{Landi:2023ktg}. In the homodyne unraveling, we consider the {\it quadrature} operator:
\begin{equation}
    x_{k} = L_{k} e^{-i \phi_{k}} + L_{k}^{\dagger} e^{i \phi_{k}}.
\end{equation}
Here, $\expval{x_{k}}$ can be measured, instead of $\expval{ L_{k}^{\dagger} L_{k}}$ as in the photodetection, with the output current.
Defining the super-operator
\begin{equation}
    {\cal{H}}\hat{\rho} =  \sum_{k} \nu_{k} \left( e^{- i \phi_{k}} L_{k} \hat{\rho} + e^{i \phi_{k}} \hat{\rho} L_{k}^{\dagger}\right),
\end{equation}
we can obtain the average homodyne current in the steady state
\begin{equation}
    J_{\rm Hom} = \mathrm{Tr}\left[ {\cal{H}} \hat{\rho}_{\rm ss} \right] = \sum_{k} \nu_{k} \expval{x_{k}}.
\end{equation}
Similarly, the corresponding correlation function in the steady state is given by
\begin{equation}
    F_{\rm Hom}(\tau) = K_{\rm Hom}\delta(\tau)+ {\rm Tr}\left[ {\cal{H}} e^{\cal{L} |\tau|}{\cal{H}}\hat{\rho}_{\rm ss} \right] - J_{\rm Hom}^{2},
\end{equation}
with 
\begin{equation}
    K_{\rm Hom} = \sum_{k} \nu_{k}^{2}.
\end{equation}
Unlike in the photodetection measurement, the white-noise intensity $K_{\rm Hom}$ is independent of the state of the system.
As in the main text, we define the correlation function as
\begin{equation}
    C_{\rm Hom}(\tau) \equiv F_{\rm Hom}(\tau) - K_{\rm Hom}\delta(\tau).
\end{equation}

\begin{figure}[htbp]
    \centering
    \includegraphics[width=0.8\linewidth]{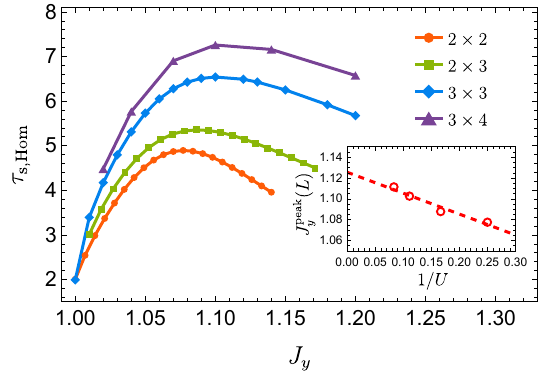}
    \caption{Plot of $\tau_{\rm s,Hom}$ as a function of $J_{y}$. The inset shows the value of $J_{y}$ at which $\tau_{\rm s,Hom}$ becomes maximum versus the inverse of the system size together with a linear fitting (dashed line).}
    \label{fig:QDplot}
\end{figure}

Here, we examine the behavior of $C_{\rm Hom}$ around the critical point within the dissipative XYZ model. We focus on the case of $\phi_{k}=0$. Fig.~\ref{fig:FplotHomo} shows the correlation function $C_{\rm Hom}(\tau)$ normalized by $C_{\rm Hom}(0)$ for several values of $J_{y}$ in the $3\times3$ lattice.

We find that the dynamic crossover in the correlation function, as observed in the photodetection, does not appear in the homodyne detection, while the plot seems to indicate the critical slowing down around the critical point. Computing the characteristic timescale defined as 
\begin{equation}
    \tau_{\rm s, Hom} \equiv \frac{1}{C_{\rm Hom}(0)}\int_{0}^{\infty} \dd \tau ~C_{\rm Hom}(\tau),\label{diffHom}
\end{equation}
we confirm that $\tau_{\rm s, Hom}$ also has a maximum at a certain value of $J_{y}$ as shown in Fig.~\ref{fig:QDplot}. However, the inset of Fig.~\ref{fig:QDplot} shows that the critical point predicted by $\tau_{\rm s, Hom}$ is much bigger than that with $\tau_{\rm s}$. This result implies that the photodetection is an appropriate measurement to characterize the criticality of DPTs compared to the homodyne detection in terms of the dynamical properties we focused on. The similar measurement-dependent criticality in the driven-dissipative Kerr model has been discussed in \cite{kewming2022diverging}. The experimental analysis to quantify the critical slowing down with the heterodyne detection has been conducted in a similar setup \cite{beaulieu2025observation}. The measurement dependence of the dynamical properties we studied in this paper should be investigated in depth, which is left for future work.

\end{document}